\documentclass[useAMS,a4paper,10pt,referee]{mn2e}
\usepackage{times}
\usepackage{graphicx}
\usepackage{amssymb}
\usepackage{longtable}

\title[Pc-scale jet in 3C~48]{Kinematics of the parsec-scale radio jet in
3C~48}

\author[An et al.]
{T. An$^{1}$\thanks{E-mail: antao@shao.ac.cn}, X.Y.~Hong$^{1}$,
M.J.~Hardcastle$^{2,3}$, D.M.~Worrall$^{3}$, T.~Venturi$^{4}$,
\newauthor
T.J.~Pearson$^{5}$, Z.-Q.~Shen$^{1}$, W.~Zhao$^{1}$ and W.X.~Feng$^{6}$ \\
$^{1}$ Shanghai Astronomical Observatory, Chinese Academy of Sciences,
200030, Shanghai, China  \\
$^{2}$ School of Physics, Astronomy and Mathematics, University of
Hertfordshire, College Lane, Hatfield AL10 9AB\\
$^{3}$ Department of Physics, University of Bristol, Tyndall Avenue,
Bristol BS8 1TL\\
$^{4}$ INAF -- Istituto di Radioastronomia, I-40129, Bologna, Italy \\
$^{5}$ California Institute of Technology, Mail Stop 105-24,
Pasadena, CA 91125, USA \\
$^{6}$ Liaocheng University, 252059, Liaocheng, China }

\begin{document}

\date{\today}

\pagerange{\pageref{firstpage}--\pageref{lastpage}} \pubyear{2009}

\maketitle
\label{firstpage}

\begin{abstract}
We present results on the compact steep-spectrum quasar 3C~48 from
observations with the Very Long Baseline Array (VLBA), the Multi-Element
Radio Linked Interferometer Network (MERLIN) and the European VLBI
Network (EVN) at multiple radio frequencies. In the 1.5-GHz VLBI images,
the radio jet is characterized by a series of bright knots. The active
nucleus is embedded in the southernmost VLBI component A, which is
further resolved into two sub-components A1 and A2 at 4.8 and 8.3 GHz.
A1 shows a flat spectrum and A2 shows a steep spectrum. The most
strongly polarized VLBI components are located at component C
$\sim$0.25 arcsec north of the core, where the jet starts to bend to the
northeast. The polarization angles at C show gradual changes across the
jet width at all observed frequencies, indicative of a gradient in the
emission-weighted intrinsic polarization angle across the jet and
possibly a systematic gradient in the rotation measure; moreover, the
percentage of polarization increases near the curvature at C, likely
consistent with the presence of a local jet-ISM interaction and/or
changing magnetic-field directions. The hot spot B shows a higher
rotation measure, and has no detected proper motion. These facts provide
some evidence for a stationary shock in the vicinity of B. Comparison of
the present VLBI observations with those made 8.43 years ago suggests a
significant northward motion for A2 with an apparent transverse velocity
$\beta_{app}=3.7\pm0.4\, c$. The apparent superluminal motion suggests
that the relativistic jet plasma moves at a velocity of $\gtrsim0.96\,
c$ if the jet is viewed at an inclination angle less than $20\degr$. A
simple precessing jet model and a hydrodynamical isothermal jet model
with helical-mode Kelvin-Helmholtz instabilities are used to fit the
oscillatory jet trajectory of 3C~48 defined by the bright knots.
\end{abstract}


\section{INTRODUCTION}\label{section:intro}

Compact Steep Spectrum (CSS) sources are a population of powerful radio
sources with projected linear size less than 20 kpc and steep high radio
frequency spectrum $\alpha<-0.5$ \footnote{In the present paper, the
spectral index is defined as $S_\nu\propto\nu^\alpha$.} (Peacock \& Wall
1982, Fanti et al. 1990, and review by O'Dea 1998 {\bf and Fanti 2009}).
Kinematical studies
of the hot spots and analysis of the high-frequency turnover in the
radio spectrum due to radiative cooling imply ages for CSS sources in
the range 10$^2$--10$^5$ yr (e.g., Owsianik, Conway \& Polatidis 1998;
Murgia et al. 1999). The sub-galactic size of CSS sources has been used
to argue that CSS sources are probably young radio sources, (the `youth'
model: Fanti et al. 1995; Readhead et al. 1996). However, another
interpretation attributes the apparent compactness of the CSS sources to
being strongly confined by the dense ISM in the host galaxy (the
`frustration' model: van Breugel, Miley \& Heckman 1984). Spectroscopic
observations of CSS sources provide evidence for abundant gas reservoirs
in the host galaxies and strong interaction between the radio sources
and the emission-line clouds \cite{ODe02}. Some CSS sources have been
observed to have high-velocity clouds (as high as $\sim500$
km\,s$^{-1}$) in the Narrow-Line Region (NLR), presumably driven by
radio jets or outflows; an example is 3C~48 \cite{Cha99,Sto07}. In
addition, many CSS sources show distorted radio structures, suggestive
of violent interaction between the jet and the ambient interstellar
medium \cite{Wil84,Fan85,Nan91,Spe91,Nan92,Aku91}. The ample supply of
cold gas in their host galaxies and their strong radio activity, which
results in a detection rate as high as $\sim30$ per cent in flux-density
limited radio source surveys \cite{Pea82,Fan90}, make CSS sources good
laboratories for the study of AGN triggering and feedback.

3C~48 ($z=0.367$) is associated with the first quasar to be discovered
\cite{Mat61,Gre63} in the optical band. Its host galaxy is brighter than
that of most other low redshift quasars. The radio source 3C~48 is
classified as a CSS source due to its small size and steep radio
spectrum \cite{Pea82}. Optical and NIR spectroscopic observations
suggest that the active nucleus is located in a gas-rich environment and
that the line-emitting gas clouds are interacting with the jet material
\cite{Can90,Sto91,Cha99,Zut04,Kri05,Sto07}. VLBI images
\cite{Wil90,Wil91,Nan91,Wor04} have revealed a disrupted jet in 3C~48,
indicative of strong interactions between the jet flow and the dense
clouds in the host galaxy. Although some authors
\cite{Wil91,Gup05,Sto07} have suggested that the vigorous radio jet is
powerful enough to drive massive clouds in the NLR at speeds up to 1000
km~s$^{-1}$, the dynamics of the 3C~48 radio jet have yet to be well
constrained. Due to the complex structure of the source, kinematical
analysis of 3C~48 through tracing proper motions of compact jet
components can only be done with VLBI observations at 4.8 GHz and higher
frequencies, but until now the required multi-epoch high-frequency VLBI
observations had not been carried out.

In order to study the kinematics of the radio jet for comparison with
the physical properties of the host galaxy, we observed 3C~48 in full
polarization mode with the VLBA at 1.5, 4.8 and 8.3 GHz in 2004, and
with the EVN and MERLIN at 1.65 GHz in 2005. Combined with earlier VLBA
and EVN observations, these data allow us to constrain the dynamics of the
jet on various scales. Our new observations and our interpretation of
the data are presented in this paper. The remainder of the paper is laid
out as follows. Section 2 describes the observations and data reduction;
Section 3 presents the total intensity images of 3C~48; and Section 4
discusses the spectral properties and the linear polarization of the
components of the radio jet. In Section 5, we discuss the implications
of our observations for the kinematics and dynamics of the radio jet.
Section 6 summarizes our results. Throughout this paper we adopt a
cosmological model with Hubble constant $H_0$=70 km~s$^{-1}$~Mpc$^{-1}$,
$\Omega_m=0.3$, and $\Omega_{\Lambda}=0.7$. Under this cosmological
model, a 1-arcsec angular separation corresponds to a projected linear
size of 5.1 kpc in the source frame at the distance of 3C 48
($z=0.367$).

\section{OBSERVATIONS AND DATA REDUCTION}

The VLBA observations (which included a single VLA antenna) of 3C~48
were carried out at 1.5, 4.8, and 8.3 GHz on 2004 June 25. The EVN and
MERLIN observations at 1.65 GHz were simultaneously made on 2005 June 7.
Table \ref{tab:obs} lists the parameters of the VLBA, EVN and MERLIN
observations. In addition to our new observations, we made use of the
VLBA observations described by Worrall et al. (2004) taken in 1996 at
1.5, 5.0. 8.4 and 15.4 GHz.

\subsection{VLBA observations and data reduction}

The total 12 hours of VLBA observing time were evenly allocated among the
three frequencies. At each frequency the effective observing time on
3C~48 is about 2.6 hours. The data were recorded at four observing
frequencies (IFs) at 1.5 GHz and at two frequencies at the other two
bands, initially split into 16 channels each, in full polarization mode.
The total bandwidth in each case was 32 MHz. The detailed data reduction
procedure was as described by Worrall et al. (2004) and was carried out
in {\sc aips}. We used models derived from our 1996 observations to
facilitate fringe fitting of the 3C~48 data. Because the source
structure of 3C~48 is heavily resolved at 4.8 and 8.3 GHz, and missing
short baselines adds noise to the image, the initial data were not
perfectly calibrated. We carried out self-calibration to further correct
the antenna-based phase and amplitude errors. This progress improves the
dynamic range in the final images.

Polarization calibration was also carried out in the standard manner.
Observations of our bandpass calibrator, 3C~345, were used to determine
the R-L phase and delay offsets. The bright calibrator source DA~193 was
observed at a range of parallactic angles and we used a model image of
this, made from the Stokes $I$ data, to solve for instrumental
polarization. Our observing run included a snapshot observation of the
strongly polarized source 3C~138. Assuming that the polarization
position angle (or the E-Vector position angle in polarization images,
`EVPA') of 3C~138 on VLBI scales at 1.5 GHz is the same as the value
measured by the VLA, and we used the measured polarization position
angle of this source to make a rotation of $94\degr$ of the position
angles in our 3C~48 data. We will show later that the corrected EVPAs of
3C~48 at 1.5 GHz are well consistent with those derived from the
1.65-GHz EVN data that are calibrated independently. At 4.8 GHz, 3C~138
shows multiple polarized components; we estimated the polarization angle
for the brightest polarized component in 3C~138 from Figure 1 in Cotton
et al. 2003, and determined a correction of $-55\degr$ for the 3C~48
data. After the rotation of the EVPAs, the polarized structures at
4.8~GHz are basically in agreement with those at 1.5 GHz. At 8.3~GHz the
polarized emission of 3C~138 is too weak to be used to correct the
absolute EVPA; we therefore did not calibrate the absolute
EVPAs at 8.3 GHz.

\subsection{EVN observations and data reduction}

The effective observing time on 3C~48 was about 8 hours. Apart from
occasional RFI (radio frequency interference), the whole observation ran
successfully. The data were recorded in four IFs. Each IF was split into
16 channels, each of 0.5-MHz channel width. In addition to 3C~48 we
observed the quasars DA~193 and 3C~138 for phase calibration. 3C 138 was
used as a fringe finder due to its high flux density of $\sim$9~Jy at
1.65 GHz.

The amplitude of the visibility data was calibrated using the system
temperatures, monitored during the observations, and gain curves of
each antenna that were measured within 2 weeks of the observations. The
parallactic angles were determined on each telescope and the data were
corrected appropriately before phase and polarization calibration. We
corrected the ionospheric Faraday rotation using archival ionospheric
model data from the CDDIS. DA~193 and OQ~208 were used to calibrate the
complex bandpass response of each antenna. We first ran fringe fitting
on DA~193 over a 10-minute time span to align the multi-band delays.
Then a full fringe fitting using all calibrators over the whole
observing time was carried out to solve for the residual delays and
phase rates. The derived gain solutions were interpolated to calibrate
the 3C~48 visibility data. The single-source data were split for hybrid
imaging. We first ran phase-only self-calibration of the 3C 48 data to
remove the antenna-based, residual phase errors. Next we ran three
iterations of both amplitude and phase self-calibration to improve the
dynamic range of the image.

DA~193 is weakly polarized at centimetre wavelengths (its fractional
polarization is no more than 1 per cent at 5 GHz, Xiang et al. 2006),
and was observed over a wide range of parallactic angles to calibrate
the feed response to polarized signals. The instrumental polarization
parameters of the antenna feeds (the so-called `D-terms') were
calculated from the DA~193 data and then used to correct the phase of
the 3C~48 data. The absolute EVPA was then calibrated from
observations of 3C~138 \cite{Cot97b,Tay00}. A comparison between the
apparent polarization angle of 3C~138 and the value from the VLA
calibrator monitoring program (i.e., $-15\degr$ at 20 cm
wavelength) leads to a differential angle $-22\degr$, which was
applied to correct the apparent orientation of the E-vector for the
3C~48 data. After correction of instrumental polarization and absolute
polarization angle, the cross-correlated 3C~48 data were used to produce
Stokes $Q$ and $U$ images, from which maps of linear polarization intensity
and position angle were produced.

\subsection{MERLIN observations and data reduction}

The MERLIN observations of 3C~48 were performed in the fake-continuum
mode: the total bandwidth of 15 MHz was split into 15 contiguous
channels, 1 MHz for each channel. A number of strong, compact
extragalactic sources were interspersed into the observations of 3C~48
to calibrate the complex antenna gains.

The MERLIN data were reduced in {\sc aips} following the standard
procedure described in the MERLIN cookbook. The flux-density scale was
determined using 3C~286 which has a flux density of 13.7 Jy at 1.65 GHz.
The phases of the data were corrected for the varying parallactic angles
on each antenna. Magnetized plasma in the ionosphere results in an
additional phase difference between the right- and left-handed signals,
owing to Faraday rotation. This time-variable Faraday rotation tends to
defocus the polarized image and to give rise to erroneous estimates of
the instrumental polarization parameters. We estimated the ionospheric
Faraday rotation on each antenna based on the model suggested in the
{\sc aips} Cookbook, and corrected the phases of the visibilities
accordingly. DA~193, OQ~208, PKS~2134+004 and 3C~138 were used to
calibrate the time- and elevation-dependent complex gains. These gain
solutions from the calibrators were interpolated to the 3C~48 data. The
calibrated data were averaged in 30-second bins for further imaging
analysis. Self-calibration in both amplitude and phase was performed to
remove residual errors.

The observations of OQ~208 were used to calculate the instrumental
polarization parameters of each antenna assuming a point-source model.
The derived parameters were then applied to the multi-source data. We
compared the right- and left-hand phase difference of the 3C~286
visibility data with the phase difference value derived from the VLA
monitoring program (i.e., $66\degr$ at 20 cm, Cotton et al. 1997b;
Taylor \& Myers 2000), and obtained a differential angle of $141\degr$.
This angle was used to rotate the EVPA of the polarized data for 3C~48.

\subsection{Combination of EVN and MERLIN data}

After self-calibration, the EVN and MERLIN data of 3C~48 were combined
to make an image with intermediate resolution and high sensitivity. The
pointing centre of the MERLIN observation was offset by 0.034 arcsec to
the West and 0.378 arcsec to the North with respect to the EVN pointing
centre (Table \ref{tab:obs}). Before combination, we first shifted the
pointing centre of the MERLIN data to align with that of the EVN data.
The Lovell and Cambridge telescopes took part in both the EVN and MERLIN
observations. We compared the amplitude of 3C~48 on the common
Lovell--Cambridge baseline in the EVN and MERLIN data, and re-scaled the
EVN visibilities by multiplying them by a factor of 1.4 to match the
MERLIN flux. After combination of EVN and MERLIN visibility data, we
performed a few iterations of amplitude and phase self-calibration to
eliminate the residual errors resulting from minor offsets in
registering the two coordinate frames and flux scales.

\section{RESULTS -- total intensity images}

\begin{figure}
\begin{center}
\includegraphics[scale=0.8,angle=0]{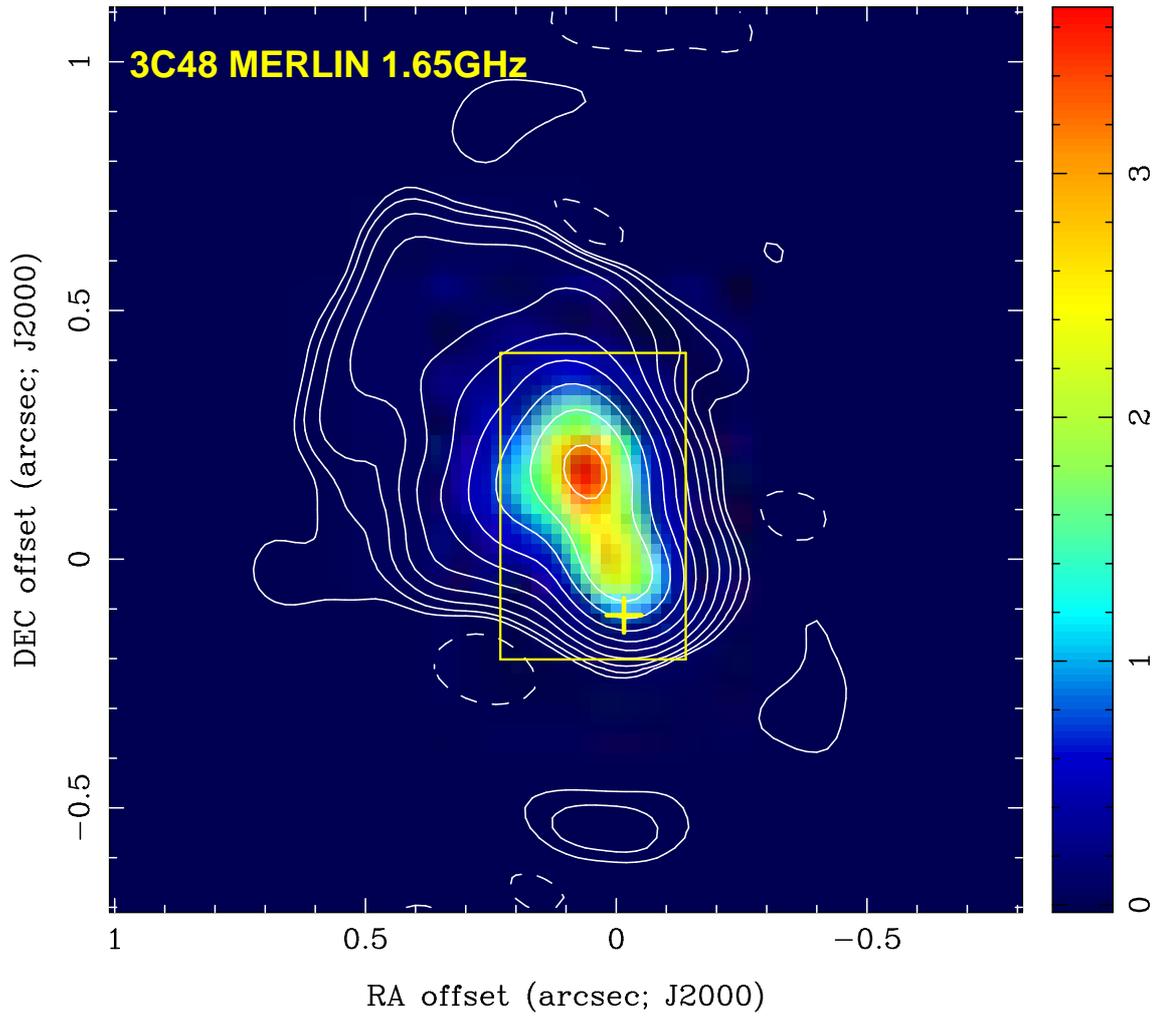}
\end{center}
\caption{Total intensity (Stokes $I$) image of 3C~48 from the MERLIN
observation at 1.65 GHz.
The image was made with uniform weighting.
The restoring beam is 138$\times$115 (mas), PA=65.1\degr.
The phase centre is at RA=01$^h$37$^d$41$^s$.29949,
Dec=$+$33\degr09\arcmin35\arcsec.1338.
The {\it r.m.s.} noise in the image measured in an
off-source region is $\sim$1.3 mJy b$^{-1}$, corresponding to a dynamic
range of $\sim$2800:1 in the image.
The contours are 6 mJy~b$^{-1}\times$(-2, 1, 2, 4, ..., 512).
The cross denotes the location of the hidden AGN.
The square marks the region in which compact jet dominates the
emission structure.
}
\label{fig:MERcont}
\end{figure}

Figures \ref{fig:MERcont} and \ref{fig:vlbimap} exhibit the total
intensity images derived from the MERLIN, VLBA and EVN data. The final
images were created using the {\sc aips} and {\sc miriad} software
packages as well as the {\sc mapplot} program in the Caltech VLBI
software package.

\subsection{MERLIN images}

Figure \ref{fig:MERcont} shows the total intensity image of 3C~48 from
the MERLIN observations. We used the multi-frequency synthesis
technique to minimize the effects of bandwidth smearing, and assumed an
optically thin synchrotron spectral index ($\alpha=-0.7$) to scale the
amplitude of the visibilities with respect to the central frequency
when averaging the data across multiple channels. The final image was
produced using a hybrid of the Clark (BGC CLEAN) and Steer (SDI CLEAN)
deconvolution algorithms. The image shows that the source structure is
characterized by two major features: a compact component contributing
about half of the total flux density (hereafter referred to as the
`compact jet'), and an extended component surrounding the compact jet
like a cocoon (hereafter called the `extended envelope'). The compact
jet is elongated in roughly the north-south direction, in alignment
with the VLBI jet. The galactic nucleus corresponding to the central
engine of 3C~48 is associated with VLBI component A
\cite{Sim90,Wil91}. It is embedded in the southern end of the compact
jet. The emission peaks at a location close to the VLBI jet component
D; the second brightest component in the compact jet is located in the
vicinity of the VLBI jet component B2 (Figure \ref{fig:vlbimap}: see
Section \ref{section:vlbimap}). The extended envelope extends out to
$\sim$1 arcsec north from the nucleus. At $\sim$0.25 arcsec north of
the nucleus, the extended component bends and diffuses toward the
northeast. The absence of short baselines ($uv<30 k\lambda$) results
in some negative features (the so-called `negative bowl' in synthesis
images) just outside the outer boundary of the envelope.

The integrated flux density over the whole source is 14.36$\pm$1.02 Jy
(very close to the single-dish measurement), suggesting that there is
not much missing flux on short spacings. The uncertainty we assign
includes both the systematic errors and the {\it r.m.s.} fluctuations in
the image. Since the calibrator of the flux density scale, 3C~286, is
resolved on baselines longer than 600 k$\lambda$ \cite{An04,Cot97a}, a
model with a set of CLEAN components was used in flux density
calibration instead of a point-source model. We further compared the
derived flux density of the phase calibrator DA~193 from our
observations with published results \cite{Sta98,Con98}. The comparison
suggests that the flux density of DA~193 from our MERLIN observation was
consistent with that from the VLBI measurements to within 7 per cent. We
note that this systematic error includes both the amplitude calibration
error of 3C~286 and the error induced by the intrinsic long-term
variability of DA~193; the latter is likely to be dominant.

The optical and NIR observations \cite{Sto91,Cha99,Zut04} detect a
secondary continuum peak, denoted 3C~48A, at $\sim$1 arcsec northeast
of the optical peak of 3C~48. Although MERLIN would be sensitive to
any compact structure with this offset from the pointing centre, we
did not find any significant radio emission associated with 3C~48A.
There is no strong feature at the position of 3C~48A even in
high-dynamic-range VLA images \cite{Bri95,Feng05}. It is possibly that
the radio emission from 3C~48A is intrinsically weak if 3C~48A is a
disrupted nucleus of the companion galaxy without an active AGN
\cite{Sto91} or 3C~48A is an active star forming region \cite{Cha99}.
In either case, the emission power of 3C~48A would be dominated by
thermal sources and any radio radiation would be highly obscured by
the surrounding interstellar medium.

\subsection{VLBA and EVN images}\label{section:vlbimap}

\begin{figure}
\begin{center}
\includegraphics[scale=0.8]{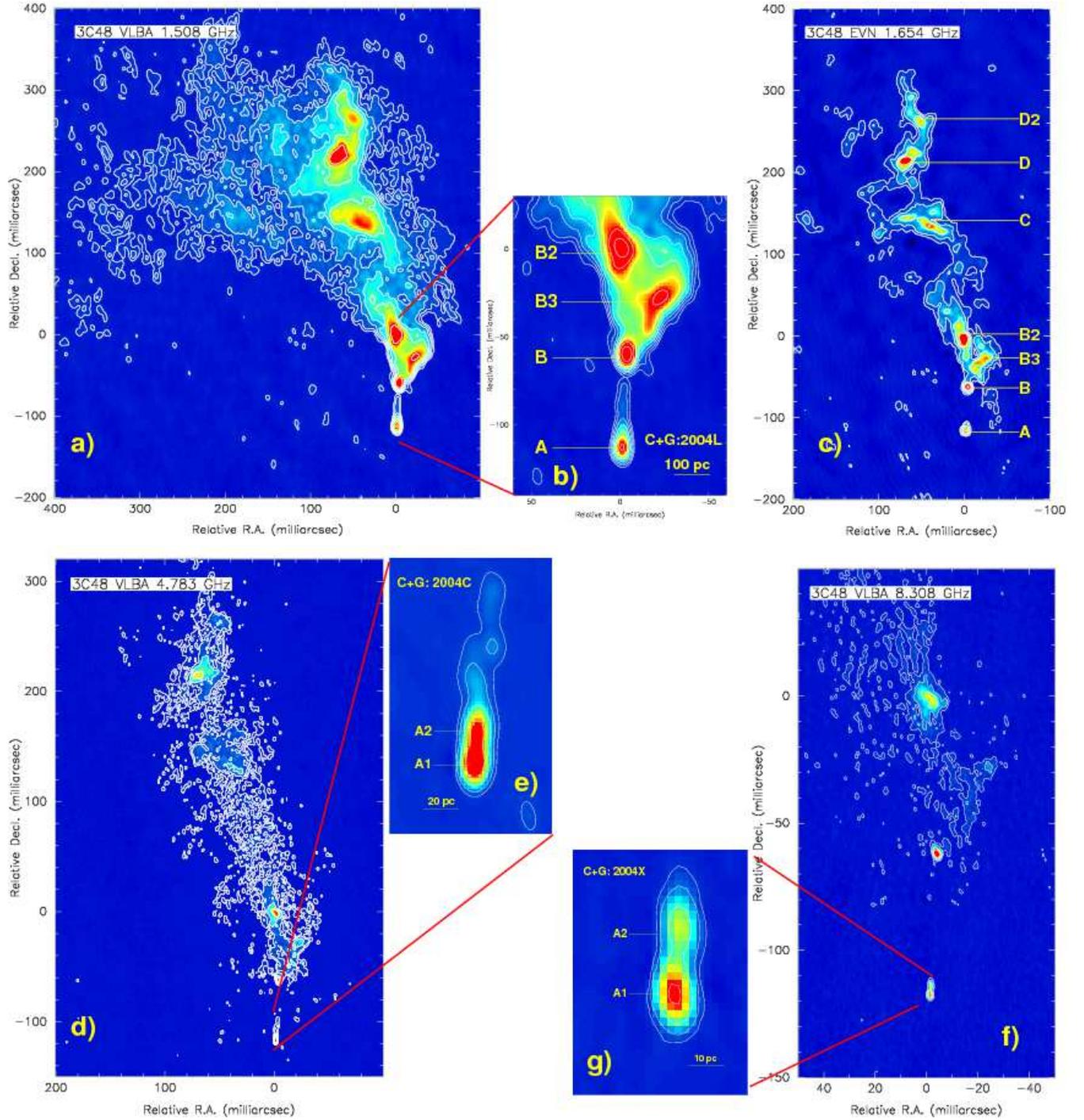}
\end{center}
\caption{Total intensity (Stokes $I$) image of 3C~48 from VLBA and EVN
observations. The phase centres of all images have been shifted to
RA=01$^h$37$^d$41$^s$.29949, Dec=$+$33\degr09\arcmin35\arcsec.1338.
Table \ref{tab:figpar} presents the image parameters. The horizon
bars in the subpanels illuminates the length scale in projection. A
number of bright components are labeled in the images. }
\label{fig:vlbimap}
\end{figure}

Figure \ref{fig:vlbimap} shows the compact radio jet of 3C~48 on various
scales derived from the VLBA and EVN observations. Table
\ref{tab:figpar} gives the parameters of the images.

The VLBI data have been averaged on all frequency channels in individual
IFs to export a single-channel dataset. The visibility amplitudes on
each IF have been corrected on the assumption of a spectral index of
$-0.7$.

The total-intensity images derived from the 1.5-GHz VLBA and 1.65-GHz
EVN data are shown in Figures \ref{fig:vlbimap}-a to
\ref{fig:vlbimap}-c. The jet morphology we see is consistent with
other published high-resolution images
\cite{Wil90,Wil91,Nan91,Wor04,Feng05}. The jet extends $\sim$0.5
arcsec in the north-south direction, and consists of a diffuse plume
in which a number of bright compact knots are embedded. We label these
knots in the image using nomenclature consistent with the previous
VLBI observations \cite{Wil91,Wor04} (we introduce the labels B3 and
D2 for faint features in the B and D regions revealed by our new
observations). The active nucleus is thought to be located at the
southern end of the jet, i.e., close to the position of component A
\cite{Sim90,Wil91}. The bright knots, other than the nuclear
component A, are thought to be associated with shocks that are created
when the jet flow passes through the dense interstellar medium in the
host galaxy \cite{Wil91,Wor04,Feng05}. Figure \ref{fig:vlbimap}-b
enlarges the inner jet region of the 3C~48, showing the structure
between A and B2. At $\sim$0.05 arcsec north away from the core A, the
jet brightens at the hot spot B. B is in fact the brightest jet
knot in the VLBI images. Earlier 1.5-GHz images (Figure 1 : Wilkinson
et al. 1991; Figure 5 : Worrall et al. 2004) show only weak flux
($\sim4\sigma$) between A and B, but in our high-dynamic-range image in
Figure \ref{fig:vlbimap}-b, a continuous jet is distinctly seen to
connect A and B. From component B, the jet curves to the northwest. At $\sim$0.1 arcsec north of the nucleus, there is a
bright component B2. After B2, the jet position angle seems to have a
significant increase, and the jet bends into a second curve with a
larger radius. At 0.25 arcsec north of the nucleus, the jet runs into
a bright knot C which is elongated in the East-West direction. Here a plume of emission turns toward the northeast. The outer
boundary of the plume feature is ill-defined in this image since its
surface brightness is dependent on the {\it r.m.s.} noise in the
image. The compact jet still keeps its northward motion from component
C, but bends into an even larger curvature. Beyond component D2, the
compact VLBI jet is too weak to be detected.

At 4.8 and 8.3 GHz, most of the extended emission is resolved out
(Figures \ref{fig:vlbimap}-d to \ref{fig:vlbimap}-g) and only a few
compact knots remain visible. Figure \ref{fig:vlbimap}-e at 4.8 GHz
highlights the core-jet structure within 150 pc ($\sim$30 mas); the
ridge line appears to oscillate from side to side. At the resolution
of this image the core A is resolved into two sub-components, which we
denote A1 and A2. Figure \ref{fig:vlbimap}-g at 8.3 GHz focuses on the
nuclear region within 50 pc ($\sim$10 mas) and clearly shows two
well-separated components. Beyond this distance the brightness of the
inner jet is below the detection threshold. This is consistent with
what was seen in the 8.4- and 15.4-GHz images from the 1996 VLBA
observations \cite{Wor04}.

\begin{figure}
\begin{center}
\includegraphics[width=\textwidth]{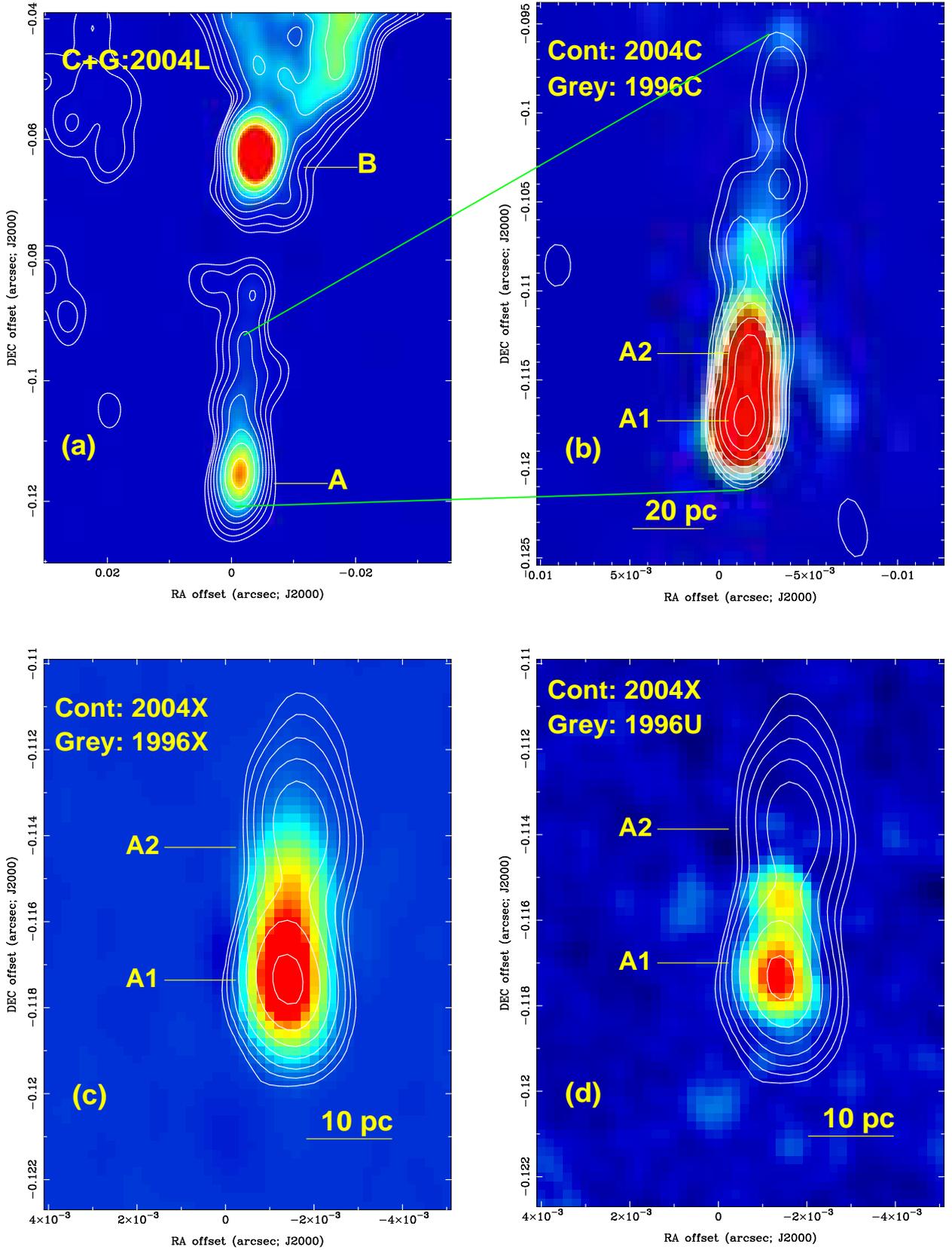}
\end{center}
\caption{Total intensity (Stokes $I$) image of 3C~48 from VLBA
observations in 2004 and 1996. {\bf top left:} 1.5 GHz image on epoch
2004. The image was made with a super-uniform weighting
 (ROBUST=$-4$ and UVBOX=3 in {\sc aips} task IMAGR)
 and restored with a beam of 6.2$\times$3.9 (mas), PA=$-30\degr$
Contours: 0.4 mJy b$^{-1}\times$ (1,2,...,128); {\bf top right:} 4.8 GHz
contours (epoch
1996) superposited on the grey-scale image (epoch 2004); {\bf bottom
left:} 8.3 GHz contours (epoch 2004) and grey-scale (epoch 1996),
{\bf bottom right:} 8.3 GHz contours (epoch 2004) and 15.4 GHz
grey-scale (epoch 1996). } \label{fig:core}
\end{figure}

Figure \ref{fig:core} focuses on the core A and inner jet out to
the hot spot B. Figure \ref{fig:core}-a shows the 1.5-GHz
image from 2004. Unlike the image already shown in Figure
\ref{fig:vlbimap}-b, this image was produced with a super-uniform
weighting of the $uv$ plane (see the caption of Figure \ref{fig:core}
for details). The high-resolution
1.5-GHz image reveals a quasi-oscillatory jet extending to a distance of
$\sim$40 mas ($\sim$200~pc) to the north of the core A. Interestingly,
Figure \ref{fig:core}-b shows similar oscillatory jet structure at
4.8-GHz on both epoch 2004 (contours) and epoch 1996 (grey-scale,
Worrall et al. 2004). The consistency of the jet morphology seen in both
1.5- and 4.8-GHz images and in both epochs may suggest that the
oscillatory pattern of the jet seen on kpc scales (Figure
\ref{fig:vlbimap}) may be traced back to the innermost jet on parsec
scales. Figure \ref{fig:core}-c shows the 8.3-GHz images in 2004
(contours) and 1996 (grey scale, Worrall et al. 2004). In 1996 (the
image denoted `1996X') the core is only slightly resolved into the two
components A1 and A2, while these are well separated by 3.5 mas (2
times the synthesized beam size) in the 2004
observations (`2004X'). Direct comparison of 1996X and 2004X images thus
provides evidence for a northward position shift of A2 between 1996 and
2004. Figure \ref{fig:core}-d overlays the 2004X contour map on the
1996U (15.4 GHz, Worrall et al. 2004) grey-scale map. Neglecting the
minor positional offset of A1 between 1996U and 2004X, possibly due to
opacity effects, this comparison of 1996U and 2004X maps is also
consistent with the idea that A2 has moved north between 1996 and 2004. We will
discuss the jet kinematics in detail in Section \ref{section:pm}.

\section{image analysis}

\subsection{Spectral index distribution along the radio jet}

\begin{figure}
\includegraphics[width=\textwidth]{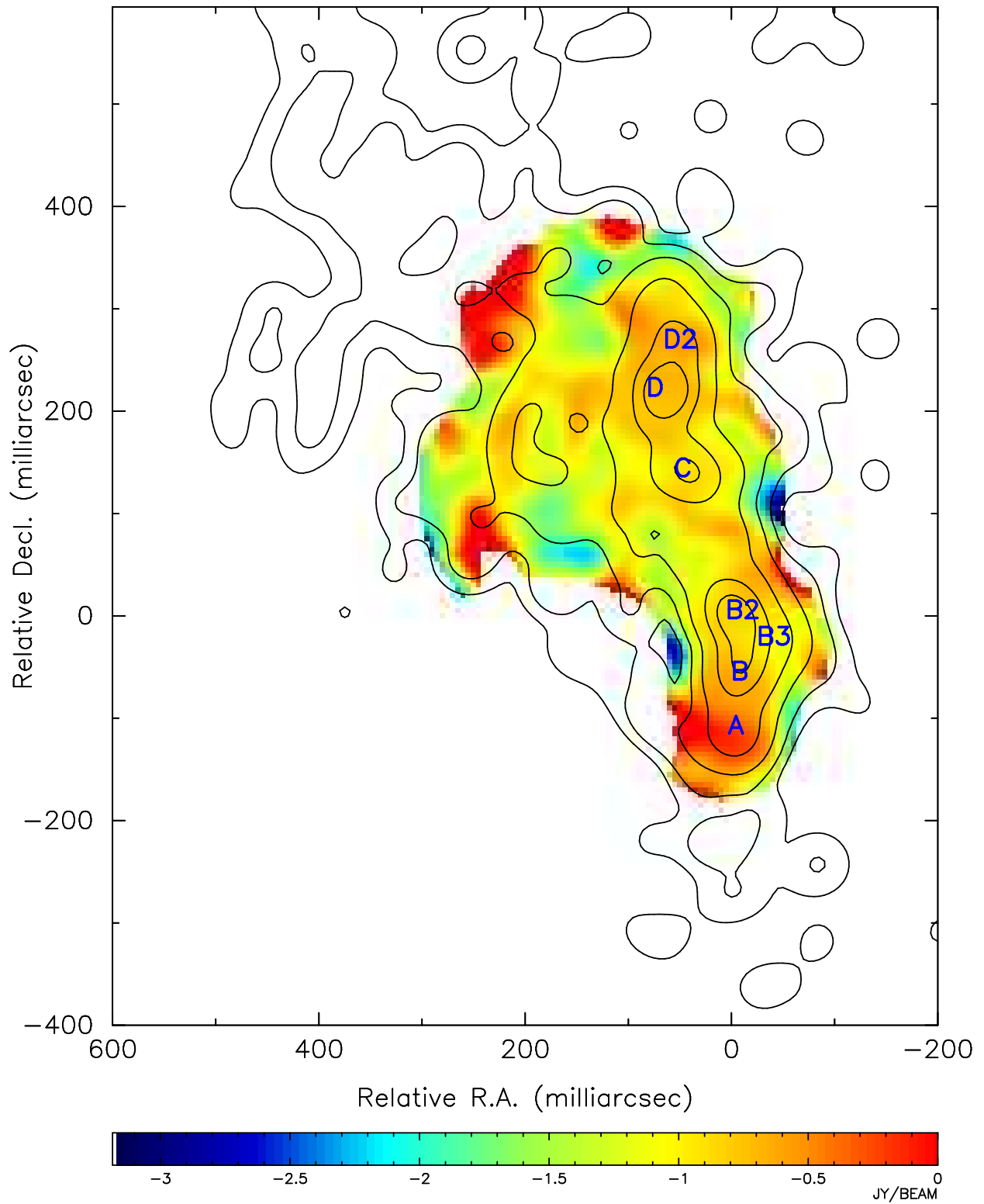}
\caption{Spectral index map (gray scale) of 3C~48 between 1.65 and
4.99 GHz. The 4.99~GHz total intensity map (contours) is derived
from the MERLIN observations on 1992 June 15 (Feng et al. 2005). The
restoring beam of the 4.99~GHz image is 40$\times$40 (mas). The
lowest contour is 0.7 mJy~b$^{-1}$, increasing in a step of 4. The
1.65 GHz data are obtained from the combined EVN and MERLIN data
observed on 2005 June 7 (the present paper). The two images are
re-produced using visibility data on the common {\it uv} range, and
restored with the same 40$\times$40 (mas) beam. Compact VLBI
components are labeled in the image. } \label{fig:spix}
\end{figure}

In order to measure the spectral properties of the 3C~48 jet, we
re-mapped the 4.99-GHz MERLIN data acquired on 1992 June 15
\cite{Feng05} and compared it with the 1.65-GHz EVN+MERLIN data
described in the present paper. The individual data sets were first
mapped with the same {\it uv} range, and convolved with the same
40$\times$40 (mas) restoring beam. Then we compared the intensities of
the two images pixel by pixel to calculate the spectral index
$\alpha^{4.99}_{1.65}$. The results are shown in Figure
\ref{fig:spix}. Component A shows a rather flat spectrum with a
spectral index $\alpha^{4.99}_{1.65}=-0.24\pm0.09$. All other bright
knots show steep spectral indices, ranging from $-0.66$ to $-0.92$.
The extended envelope in general has an even steeper spectrum with
$\alpha\lesssim-1.10$. Spectral steepening in radio sources is a
signature of a less efficient acceleration mechanism and/or the
depletion of high-energy electrons through synchrotron/Compton
radiation losses and adiabatic losses as a result of the expansion of
the plasma as it flows away from active acceleration region. The
different spectral index distribution seen in the compact jet and
extended envelope may indicate that there are different electron
populations in these two components, with the extended component
arising from an aged electron population.

\subsection{Linear polarization images}

\subsubsection{MERLIN images}
\label{section:rm-reference}

\begin{figure}
\begin{center}
\includegraphics[width=\textwidth]{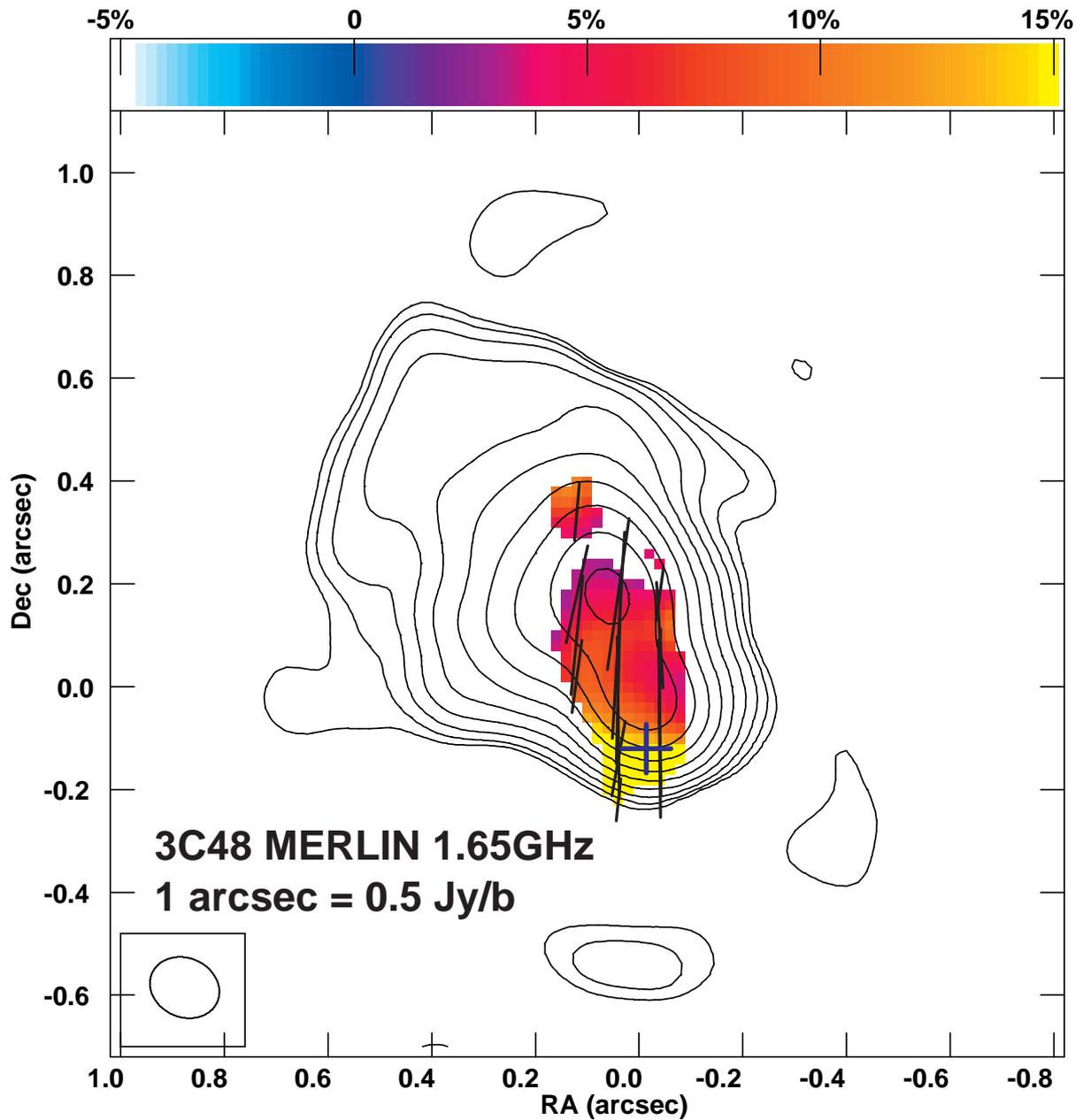}
\end{center}
\caption{Polarization structure of 3C~48 from the MERLIN observations at
1.65 GHz. The contours map is Stokes $I$ image (Figure~\ref{fig:MERcont}).
The polarization image is derived from Stokes $Q$ and $U$ images above a
4$\sigma$ cutoff (1$\sigma$=6 mJy b$^{-1}$). The wedge at the top
indicates the percentage of the polarization. The length of the bars
represents the strength of polarized emission, 1 arcsec represents 0.5
Jy b$^{-1}$. The orientation of the bar indicates the RM-corrected EVPA.
} \label{fig:MERpol}
\end{figure}

Figure \ref{fig:MERpol} displays the polarization image made from the
MERLIN data.

The majority of the polarized emission is detected in the inner region
of the source, in alignment with the compact jet. The polarized
intensity peaks in two locations. The brightest one is near the VLBI
jet component C, with an integrated polarized intensity of 0.31 Jy and
a mean percentage of polarization (defined as $\frac{\Sigma
  \sqrt{Q_i^2+U_i^2}}{\Sigma I_i}$, where $i$ represents the $i$th
polarized sub-components) of $m=5.8$ per cent. The secondary one is
located between VLBI jet components B and B2, with an integrated
polarized intensity of 0.23 Jy and a mean degree of polarization
$m=9.5$ per cent. Both of the two peaks show clear deviations from the
total intensity peaks in Figure \ref{fig:MERcont}. These measurements
of polarization structure and fractional polarization are in good
agreement with those observed with the VLA at 2-cm wavelength with a
similar angular resolution \cite{Bre84}. The integrated polarized flux
density in the whole source is 0.64$\pm$0.05 Jy and the integrated
fractional polarization is (4.9$\pm$0.4) per cent. Since the
integrated polarized intensity is in fact a vector sum of different
polarized sub-components, the percentage polarization calculated in
this way represents a lower limit. We can see from the image (Figure
\ref{fig:MERpol}) that the percentage of the polarization at
individual pixels is higher than 5 per cent, and increases toward the
south of the nucleus. A maximum value of $m\gtrsim30$ per cent is
detected at $\sim$0.045 arcsec south of the nucleus. The fractional
polarization ($m>4.9$ per cent) measured from our MERLIN observation
at 18 cm is at least an order of magnitude higher than the VLA
measurement at 20 cm, although it is consistent with the values
measured by the VLA at 6 cm and shorter wavelengths. This difference
in the fractional polarization at these very similar wavelengths is
most likely to be an observational effect due to beam depolarization,
rather than being due to intrinsic variations in the Faraday depth
(R.~Perley, private communication).

The averaged polarization angle (or EVPA) is $-18\degr\pm5\degr$ in the
polarization structure. On the basis of the new measurements of the Rotation
Measure (RM) towards 3C~48 by Mantovani et al. (2009), i.e.,
RM=$-64$ rad~m$^{-2}$ and intrinsic position angle $\phi_0=116\degr$
\cite{Sim81,Man09}, we get a polarization angle of $-4\degr$ at
1.65 GHz. This result suggests that the absolute EVPA calibration of
3C~48 agrees with the RM-corrected EVPA within 3$\sigma$. We show
in Figure \ref{fig:MERpol} the RM-corrected EVPAs. The EVPAs are well
aligned in the North-South direction, indicating an ordered magnetic
field in the Faraday screen.

\subsubsection{EVN and VLBA images}

\begin{figure}
\includegraphics[width=0.45\textwidth]{Fig6a.ps}
\includegraphics[width=0.45\textwidth]{Fig6b.ps} \\
\includegraphics[width=0.45\textwidth]{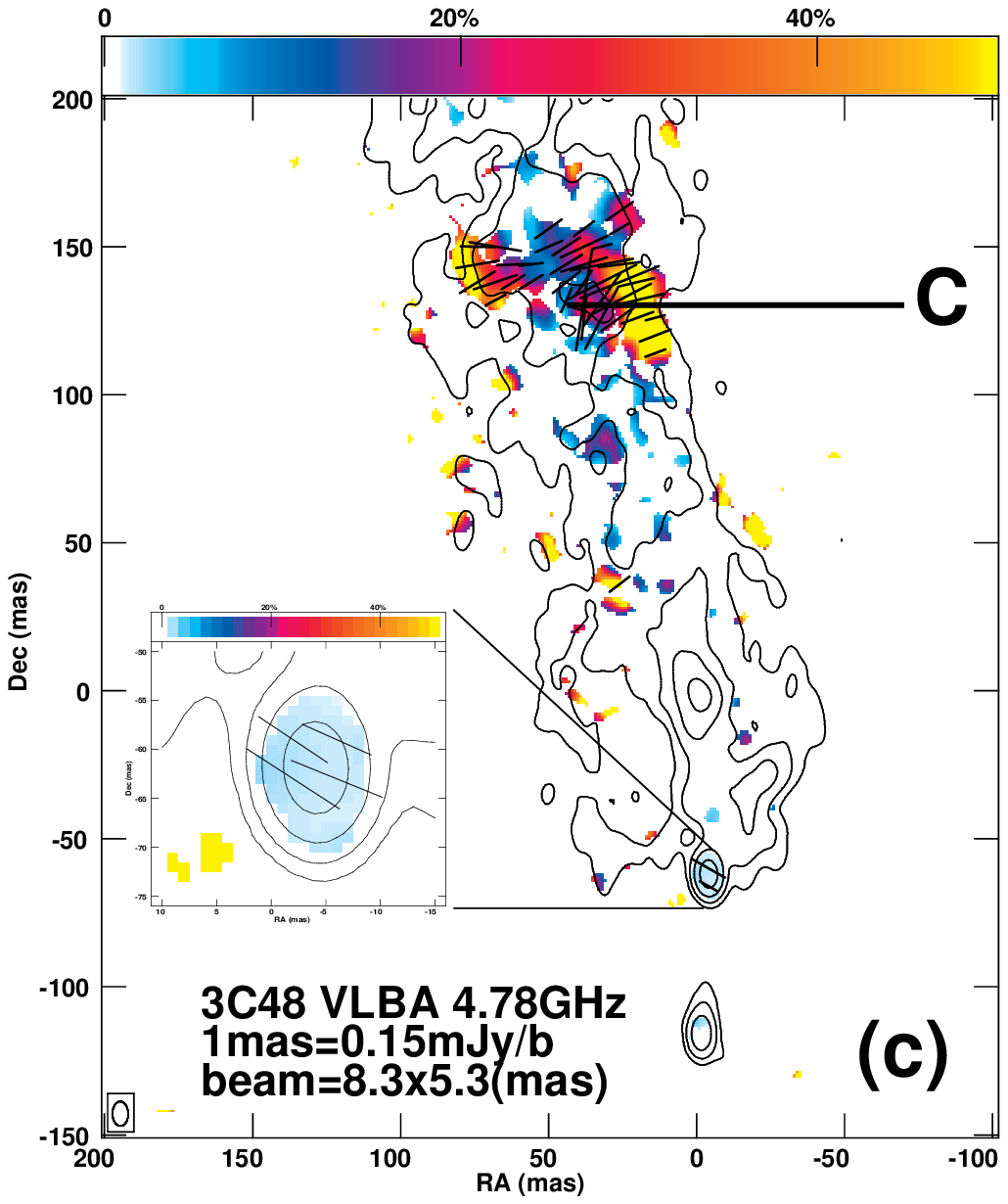}
\includegraphics[width=0.45\textwidth]{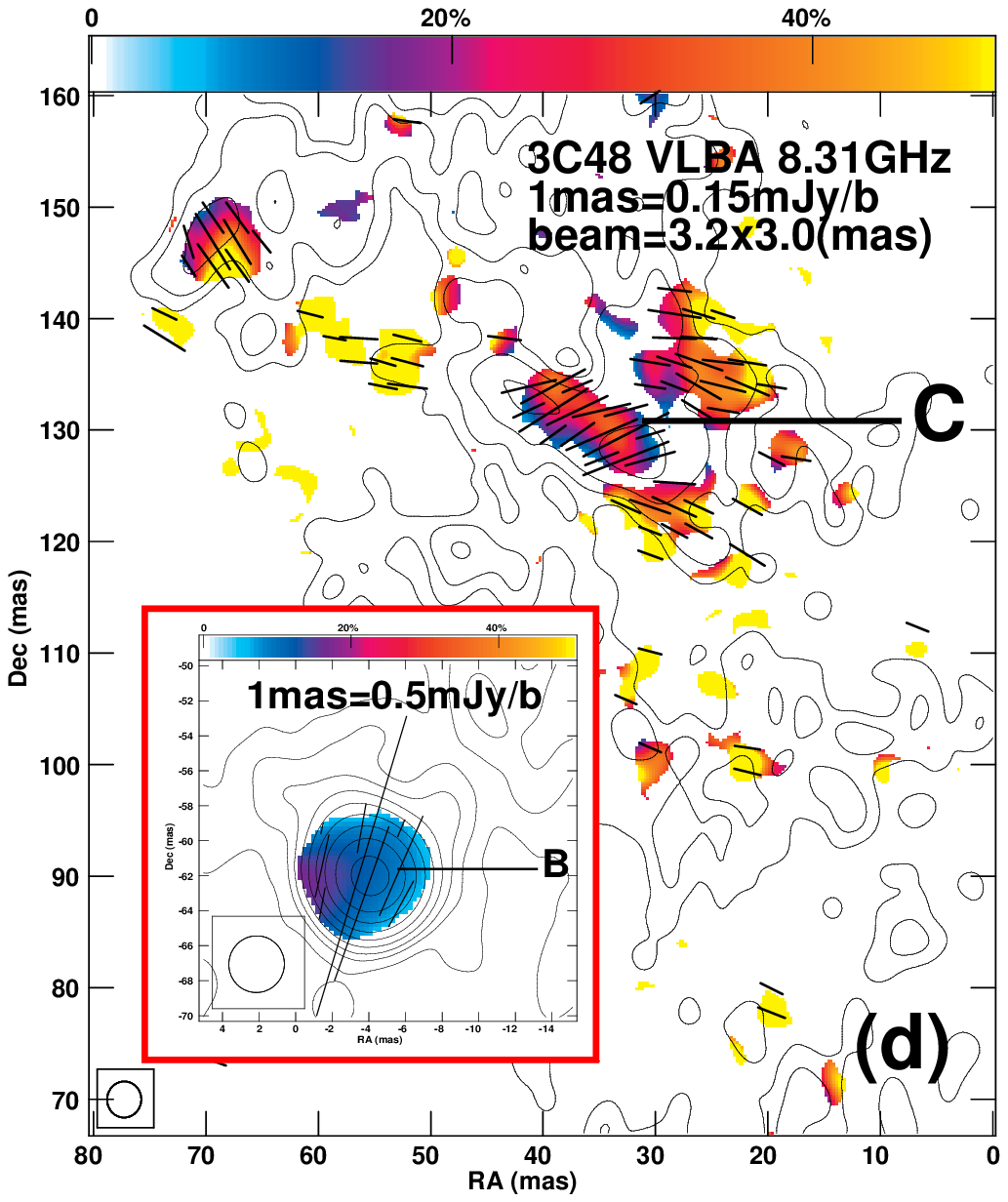}
\caption{
Polarization structure of 3C~48 derived from the EVN and VLBA data. The
contours show the total intensity (Stoke I) emission, and the grey scale
indicates the fractional polarization. The length of the bars indicates
the strength of the linear polarization intensity, and the orientation
of the bars indicates the polarization angle, which has been corrected
by the RM on the basis of the measurements by Mantovani et al. (2009).
We should note that the VLBI images show more complex polarization
structure than that shown in MERLIN image (Figure \ref{fig:MERpol}):
quantitative calculations (Figure \ref{fig:RM}) show that the RMs in the
component C region is about 1.4 times the value measured from the
overall source; moreover, the intrinsic polarization angles rotate by
$\sim 60\degr$ from the northwest edge of component C to the
southeast edge. Therefore the correction based on the overall-source RM
might not be sufficiently accurate to all sub-components, while this
uncertainty tends to small toward the higher frequencies.
{\bf (a)}: the EVN image at 1.65 GHz. Contours are 4~mJy b$^{-1}\times$
(1,4,16,64,256).
The Stokes $Q$ and $U$ maps were convolved with a 20-mas circular beam,
and we used intensities above 4$\sigma$ to calculate the polarized
intensity and polarization angle;
{\bf (b)}: contours : 1~mJy b$^{-1}\times$ (1,4,16,64,256);
{\bf (c)}: contours : 1~mJy b$^{-1}\times$ (1,4,16,64,256);
{\bf (d)}: contours : 1.2~mJy b$^{-1}\times$ (4,16,64,256);
the contours in the inset panel are 1.2~mJy b$^{-1}\times$
(1,4,8,16,32,64,128,256,512).
}
\label{fig:VLBIpol}
\end{figure}

At the resolution of the EVN, most of the polarized emission from
extended structures is resolved out. In order to map the polarized
emission with modest sensitivity and resolution, we created Stokes $Q$
and $U$ maps using only the European baselines. Figure
\ref{fig:VLBIpol}-a shows the linear polarization of 3C~48 from the
1.65 GHz EVN data. The polarized emission peaks at two components to
the East (hereafter, `C-East') and West (hereafter, `C-West') of
component C. The integrated polarized flux density is 24.8 mJy in
`C-West' and of 22.9 mJy in `C-East', and the mean percentage
polarization in the two regions is 6.3 per cent and 10.7 per cent
respectively. The real fractional polarization at individual pixels is
much higher, for the reasons discussed above (Section 4.2.1). There is
clear evidence for the existence of sub-components in `C-West' and
`C-East'; these polarized sub-components show a variety of EVPAs, and
have much higher fractional polarization than the `mean' value. The
polarization is as high as 40 per cent at the inner edge of the knot
C, which would be consistent with the existence of a shear layer
produced by the jet-ISM interaction and/or a helical magnetic field
(3C~43: Cotton et al. 2003; 3C~120: G\'{o}mez et al. 2008). Component
B, the brightest VLBI component, however, is weakly polarized with an
intensity $<$4.0 mJy~beam$^{-1}$ (percentage polarization less than
1 per cent). The nucleus A shows no obvious polarization.

The 20-cm VLBA observations were carried out in four 8-MHz bands,
centred at 1404.5, 1412.5, 1604.5 and 1612.5 MHz. In order to compare
with the 1.65-GHz EVN polarization image, we made a VLBA polarization
image (Figure \ref{fig:VLBIpol}-b) using data in the latter two bands.
This image displays a polarization structure in excellent agreement
with that detected at 1.65 GHz with the EVN, although the angular
resolution is 3 times higher than the latter: the polarized emission
mostly comes from the vicinity of component C and the fractional
polarization increases where the jet bends; the hot spot B and the
core A are weakly polarized or not detected in polarization. The 1.65-
and 1.61-GHz images show detailed polarized structure in the component-C region on a spatial scale of tens of parsecs: the polarization angle
(EVPA) shows a gradual increase across component C, with a total range
of $160\degr$, and the percentage of polarization gradually increases
from 5 per cent to $\gtrsim$30 per cent from the Western edge to the
Eastern edge at both `C-West' and `C-East'.

Figure \ref{fig:VLBIpol}-c and \ref{fig:VLBIpol}-d show the 4.8- and
8.3-GHz polarization images made with the VLBA data. Both images were made
by tapering the visibility data using a Gaussian function in order to
increase the signal-to-noise of the low-surface brightness emission.
Similar to what is seen in the 1.65 and 1.61-GHz images, component
`C-West' shows a polarization angle that increases by $80\degr$ across
the component, but these images show the opposite sense of change of
fractional polarization -- fractional polarization decreases from 60
per cent down to 20 per cent from the northwest to the southeast.
Another distinct difference is that hot spot B shows increasing
fractional polarization toward the higher frequencies, $m \sim2.0$ per
cent at 4.8 GHz and $m\sim12$ per cent at 8.3 GHz in contrast with
$m\lesssim1$ per cent at 1.6 GHz. The difference in the fractional
polarizations of B at 1.6/4.8 GHz and 8.3 GHz imply that a component
of the Faraday screen is unresolved at 1.6 and 4.8 GHz and/or that
some internal depolarization is at work. The non-detection of
polarization from the core A at all four frequencies may suggest a
tangled magnetic field at the base of the jet.

\subsection{EVPA gradient at component C and RM distribution}

We found at all four frequencies that the polarization angles undergo
a rotation by $\gtrsim 80\degr$ across the jet ridge line at both the
`C-East' and `C-West' components. There are four possible factors that
may affect the observed polarization angle: (1) the calibration of the
absolute EVPAs; (2) Faraday rotation caused by Galactic ionized gas;
(3) Faraday rotation due to gas within the 3C~48 system and (4)
intrinsic polarization structure changes. The correction of absolute
EVPAs applies to all polarization structure, so it can not explain the
position-dependent polarization angle changes at component C; in any
case, the fact that we see similar patterns at four different
frequencies, calibrated following independent procedures, rules out
the possibility of calibration error. Galactic Faraday rotation is
non-negligible (Section \ref{section:rm-reference}; $-64$ rad m$^{-2}$
implies rotations from the true position angle of $168\degr$ at 1.4
GHz, $129\degr$ at 1.6 GHz, $14.3\degr$ at 4.8 GHz and $4.8\degr$ at
8.3 GHz), and means that we expect significant differences between the
EVPA measured at our different frequencies; however, the Galactic
Faraday screen should vary on much larger angular scales than we
observe. Only factors (3) and (4), which reflect the situation
internal to the 3C~48 system itself, will give rise to a
position-dependent rotation of the EVPAs. The EVPA gradient is related
to the gradient of the RM and the intrinsic polarization angle by:
$\frac{{\rm d}\phi}{{\rm d}x}=\lambda^2\frac{{\rm d}(RM)}{{\rm
    d}x}+\frac{{\rm d}\phi_0}{{\rm d}x}$, where the first term
represents the RM gradient and the latter term represents the
intrinsic polarization angle gradient. If the systematic gradient of
EVPAs, $\frac{{\rm d}\phi}{{\rm d}x}$, were solely attributed to an RM
gradient, then $\frac{d\phi}{dx}$ would show a strong frequency
dependence; on the other hand, if $\frac{{\rm d}\phi}{{\rm d}x}$ is
associated with the change of the intrinsic polarization angle, there
is no frequency-dependence. We compared the $\frac{{\rm d}\phi}{{\rm
    d}x}$ at 1.6 and 4.8 GHz and found a ratio $\frac{{\rm d}\phi/{\rm
    d}x (1.6GHz)}{{\rm d}\phi/{\rm d}x(4.8GHz)}=1.8$. This number
falls between 1.0 (the value expected if there were no RM gradient)
and 8.8 (the ratio of $\lambda^2$), suggesting that a combination of
RM and intrinsic polarization angle gradients are responsible for the
systematic gradient of EVPAs at C. Accordingly, it is worthwhile to
attempt to measure the RM in the VLBI components of 3C~48.

The first two bands of the 20-cm VLBA data (centre frequency 1.408
GHz) are separated from the last two bands (centre frequency 1.608
GHz) by 200 MHz, indicating a differential polarization angle of $\sim
40\degr$ across the passband. The low integrated rotation measure
means that the effects of Faraday rotation are not significant
($<10\degr$) between 4.8 and 8.3 GHz, while the absolute EVPA
calibration at 8.3 GHz is uncertain; moreover, the {\it uv} sampling
at 8.3 GHz is too sparse to allow us to image identical source
structure at 1.5 and 4.8 GHz. Therefore we used the 1.408, 1.608 and
4.78 GHz data to map the RM distribution in 3C~48.

\begin{figure}
\begin{center}
\includegraphics[width=\textwidth]{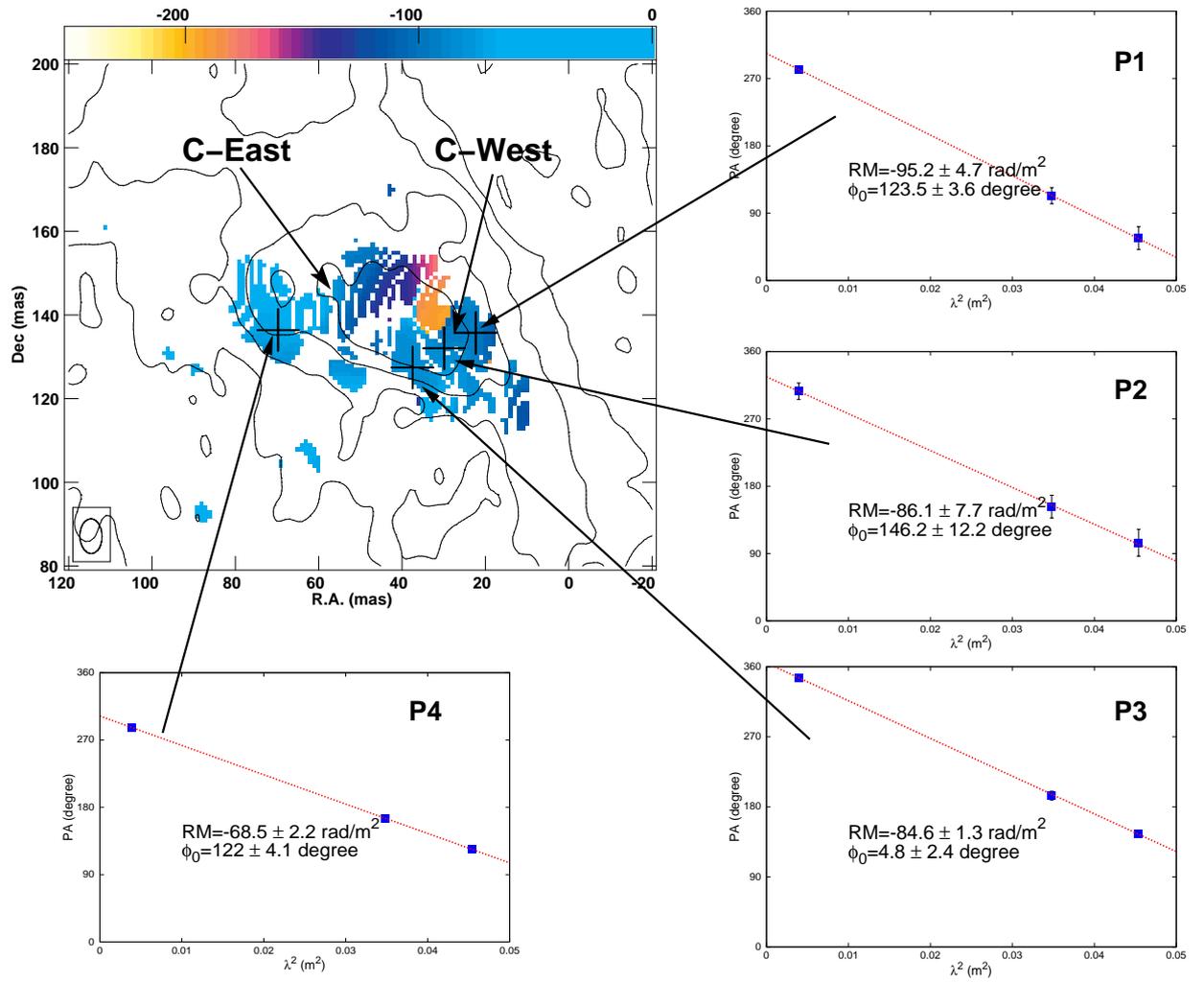}
\end{center}
\caption{RM distribution in the component-C region.
The patchy morphology is because at some pixels polarization was not
detected at all three frequencies simultaneously.
The contours represent the total intensity: 1.0 mJy
b$^{-1}\times$(1,4,8,16). The wedge at the top indicates the RM in the
observer's frame, in unit of rad m$^{-2}$. The insets show the measured
values of the observed polarization angle for four selective locations
as a function of $\lambda^2$ along with a linear fitting of rotation
measure.}
\label{fig:RM}
\end{figure}

We first re-imaged the Stokes $Q$ and $U$ data at the three
frequencies with a common {\it uv} cutoff at $>$400 k$\lambda$ and
restored with the same convolving beam. We tapered the $uv$ plane
weights when imaging the 4.78-GHz data
in order to achieve a similar intrinsic resolution to that of the
images at the two lower frequencies. We then made polarization angle
images from the Stokes $Q$ and $U$ maps. The three polarization angle
images were assembled to calculate the RM (using {\sc aips} task RM). The
resulting RM image is shown in Figure \ref{fig:RM}. The image shows a
smooth distribution of RM in the component-C region except for a
region northeast of `C-West'. The superposed plots present the
fits to the RM and intrinsic polarization angle ($\phi_0$, the
orientation of polarization extrapolated at $\lambda=0$) at four
selected locations. The polarization position angles at individual
frequencies have multiples of $\pi$ added or subtracted to remove the
$n\pi$ amibiguity. The errors in the calculated RMs and $\phi_0$ are
derived from the linear fits. We note that the systematic error due
to the absolute EVPA calibration feeds into the error on the observed
polarization angle. All four fits show a good match with a $\lambda^2$
law. The fitted parameters at `P4' in the `C-East' region are
consistent with those derived from the single-dish measurements for
the overall source \cite{Man09}. The western component (`C-West')
shows a gradient of RM from $-95$ rad m$^{-2}$ at `P1' to $-85$ rad
m$^{-2}$ at `P3', and the intrinsic polarization angle varies from
$123\degr$ (or $-57\degr$) at `P1', through $146\degr$ (or $-34\degr$)
at `P2' to $5\degr$ at `P3'. This result is in good agreement with the
qualitative analysis of the EVPA gradients above. A straightforward
interpretation of the gradients of the RMs and the intrinsic
polarization angles is that the magnetic field orientation gradually
varies across the jet ridge line; for example, a helical magnetic
field surrounding the jet might have this effect. An alternative
interpretation for the enhancement of the rotation measurement at the
edge of the jet is that it is associated with thermal electrons in a
milliarcsec-scale Faraday screens surrounding or inside the jet due to
jet-ISM interactions \cite{Cot03,Gom08}. More observations are needed
to investigate the origins of the varying RM and $\phi_0$.

The hot spot B shows a much larger difference of EVPAs between
4.8 and 8.3 GHz than is seen in component C. This
might be a signature of different RMs at B and C. A rough
calculation suggests a RM of $-330\pm60$ rad~m$^{-2}$ at B. The
high rotation measure and high fractional polarization (Section 4.2) is
indicative of a strong, ordered magnetic field in the vicinity of B. This might
be expected in a region containing a shock in which the
line-of-sight component of the magnetic field and/or the density of
thermal electrons are enhanced; in fact, the proper motion of B (Section
4.5) does provide some evidence for a stationary shock in this region.

\subsection{Physical properties of compact components in VLBI images}

In order to make a quantitative study of the radiation properties of
the compact VLBI components in 3C~48, we fitted the images
of compact components in the VLBI images from our new observations and
from the VLBA data taken in 1996 \cite{Wor04} with Gaussian models.
Measurements from the 1996 data used mapping parameters consistent with
those for the 2004 images. Table \ref{tab:model} lists the fitted
parameters of bright VLBI components in ascending frequency order.

The discrete compact components in the 4.8- and 8.3-GHz VLBA images are
well fitted with Gaussian models along with a zero-level base and slope
accounting for the extended background structure. The fit to extended
emission structure is sensitive to the {\it uv} sampling and the
sensitivity of the image. We have re-imaged the 1.5-GHz VLBA image using
the same parameters as for the 1.65-GHz EVN image, i.e., the same {\it
uv} range and restoring beam. At 1.5 and 1.65 GHz, Gaussian models are
good approximations to the emission structure of compact sources with
high signal-to-noise ratio, such as components A and B. For extended
sources (i.e., components B2 to D2) whose emission structures
are either not well modelled by Gaussian distribution, or blended with
many sub-components, model fitting with a single Gaussian model gives a
larger uncertainty for the fitted parameters. In particular, the
determination of the integrated flux density is very sensitive to the
apparent source size.

The uncertainties for the fitted parameters in Table \ref{tab:model}
are derived from the output of the {\sc aips} task {\sc JMFIT}. These
fitting errors are sensitive to the intensity fluctuations in the
images and source shapes. In most cases, the fitting errors for the
peak intensities of Gaussian components are roughly equal to the {\it
  r.m.s.} noise. We note that the uncertainty on the integrated flux
density should also contain systematic calibration errors propagated
from the amplitude calibration of the visibility data, in addition to
the fitting errors. The calibration error normally dominates over the
fitting error. The amplitude calibration for the VLBI antennas was
made from the measurements of system temperature ($T_{sys}$) at
two-minute intervals during the observations combined with the antenna
gain curves measured at each VLBI station. For the VLBA data, this
calibration has an accuracy $\lesssim$5 per cent of the amplitude
scale\footnote{See the online VLBA status summary at
  http://www.vlba.nrao.edu/astro/obstatus/current/obssum.html .}.
Because of the diversity of the antenna performance of the EVN
elements, we adopted an averaged amplitude calibration uncertainty of
5 per cent for the EVN data.

The positions of the VLBI core A1 at 4.8, 8.3 and 15.4 GHz show good
alignment within 0.4~mas at different frequencies and epochs. The
positions of the unresolved core A at 1.5 and 1.65 GHz show a systematic
northward offset by 2--4 mas relative to the position of A1 at higher
frequencies. Due to the low resolution and high opacity at 1.5 GHz, the
position of A at this frequency reflects the centroid of the blended
emission structure of the active galactic nucleus and inner 40-pc jet.
The parameters that we have derived for the compact components A, B and
B2 in epoch 1996 are in good agreement with those determined by Worrall
et al. (2004) at the same frequency band. The results for fitting to
extended knots at 1.5 and 1.65 GHz are in less good agreement. This is
probably because of the different {\it uv} sampling on short spacings,
meaning that the VLBA and EVN data sample different extended structures
in the emission.

The integrated flux densities of the VLBI components A1 and A2 in 1996X
(8.3~GHz) are higher than those in 2004X (8.3~GHz) by $\sim$100 per cent
(A1) and $\sim$60 per cent (A2), respectively. The large discrepancy in
the flux densities of A1 and A2 between epochs 1996X and 2004X can not
easily be interpreted as an amplitude calibration error of larger than
60 per cent since we do not see a variation at a comparable level
in the flux densities of components B, B2 and D. Although the {\it total} flux
densities of CSS sources in general exhibit no violent variability at radio
wavelengths, the possibility of small-amplitude ($\lesssim$100 per cent)
variability in the VLBI core and inner jet components is not ruled out.

Component A1 has a flat spectrum with
$\alpha^{8.3}_{4.8}=-0.34\pm0.04$ between 4.8 and 8.3 GHz in epoch
2004; component A2 has a rather steeper spectrum with
$\alpha^{8.3}_{4.8}=-1.29\pm0.16$ (epoch 2004). The spectral
properties of these two components support the idea that A1 is
associated with the active nucleus and suffers from synchrotron
self-absorption at centimetre radio wavelengths; in this picture, A2
is the innermost jet. The spectral indices of components B and B2
in epoch 2004 are $\alpha^{8.3}_{4.8}=-0.82\pm0.10$ (B) and
$\alpha^{8.3}_{4.8}=-0.79\pm0.10$ (B2), respectively. This is
consistent with the measurements from the 1.65 and 4.99 GHz images
(Figure \ref{fig:spix}). Component D shows a relatively flatter
spectrum in epoch 2004 with $\alpha^{8.3}_{4.8}=-0.46\pm0.06$, in
contrast to the other jet knots. While this spectral index is
consistent with those of the shock-accelerated hot spots in radio
galaxies, the flattening of the spectrum in D might also arise from a
local compression of particles and magnetic field.

Table \ref{tab:tb} lists the brightness temperatures ($T_b$) of the
compact VLBI components A1, A2 and B. All these VLBI components have a
brightness temperatures ($T_b$) higher than $10^8$K, confirming their
non-thermal origin. These brightness temperatures are well below the
$10^{11-12}$~K upper limit constrained by the inverse Compton
catastrophic \cite{KP69}, suggesting that the relativistic jet plasma is
only mildly beamed toward the line of sight. The $T_b$ of A1 is about 3
times higher than that of A2 at 4.8 and 8.3~GHz, and the $T_b$ of A1
decreases toward higher frequencies. Together with the flat spectrum and
variability of A1, the observed results are consistent with A1 being the
self-absorbed core harbouring the AGN. $T_b$ is much higher in 1996X than
2004X for both A1 and A2, a consequence of the measured flux density
variation between the two epochs.

\subsection{Proper motions of VLBI components}\label{section:pm}

The Gaussian fitting results presented in Table \ref{tab:model} may be
used to calculate the proper motions of VLBI components. In order to
search for proper motions in 3C~48, maps at different epochs
should be aligned at a compact component such as the core
\cite{Wor04}. However, thanks to our new VLBI observations we know
that aligning the cores at 1.5-GHz is not likely to be practical,
since the core structure appears to be changing on the relevant
timescales. Even at 4.8 GHz, the core still blends with the inner
jet A2 in epoch 1996C (Figure \ref{fig:core}). In contrast to these
two lower frequencies, the 8.3-GHz images have higher resolution,
better separation of A1 and A2, and less contamination from
extended emission. These make 8.3-GHz images the best choice for the
proper motion analysis. In the following discussion of proper motion
measurements we rely on the 8.3~GHz images.

We have already commented on the shift of the peak of A2 to the north
from epochs 1996X to 2004X in Figure \ref{fig:core}. A quantitative
calculation based on the model fitting results gives a positional
variation of 1.38 mas to the North and 0.15 mas to the West during a
time span 8.43 yr, assuming that the core A1 is stationary. That corresponds to a
proper motion of $\mu_\alpha = -0.018\pm0.007$ mas yr$^{-1}$ ('minus'
mean moving to the West) and $\mu_\delta=0.164\pm0.015$ mas yr$^{-1}$,
corresponding to an apparent transverse velocity of $v_\alpha =
-0.40\pm0.16 \,c$ and $v_\delta=3.74\pm0.35 \,c$. The error quoted
here includes
both the positional uncertainty derived from Gaussian fitting and the
relative offset of the reference point ({\it i.e.}, A1). That means that
we detect a significant ($>10\sigma$) proper motion for A2 moving to the
north. The apparent transverse velocity for A2 is similar to
velocities derived from other CSS and GPS sources in which
apparent superluminal motions in the pc-scale jet have been detected, e.g.,
3.3--9.7$c$ in 3C~138 \cite{Cot97b,She01}.

We also searched for evidence for proper motions of the other jet knots.
The proper motion measurement is limited by the accuracy of the
reference point alignment, our ability to make a high-precision position
determination at each epoch, and the contamination from extended
structure. We found only a
$3\sigma$ proper motion from B2, which shows a position change of
$\Delta\alpha=0.22\pm0.07$ mas and $\Delta\delta=0.48\pm0.13$ mas in
8.43 yr, corresponding to an apparent velocity of
$\beta_{app}=1.43\pm0.33\,c$ to the northeast. The measurements of the
position variation of the hot spot B between 1996X and 2004X show no
evidence for proper motion with $\mu_\alpha = 0.012\pm0.007$ mas
yr$^{-1}$ and $\mu_\delta=0.005\pm0.015$ mas yr$^{-1}$. Worrall et al.
(2004) earlier reported a $3\sigma$ proper motion for B by comparing the
the 1.5-GHz VLBA image taken in 1996 with Wilkinson et al's 1.6-GHz
image from
11.8 years previously. However, as mentioned above, the 1.5-GHz
measurements are subject to the problems of lower angular resolution,
poor reference point alignment and contamination from structural
variation. In particular, if we extrapolate the observed angular motion
of A2 back, the creation of jet component A2 took place in 1984,
therefore in 1996 A2 would still have been blended with A1 in the
1.5-GHz image within $\frac{1}{4}$ beam. The fitting of a Gaussian to
the combination of A1 and A2 at 1.5 GHz on epoch 1996 would then have
suffered from the effects of the structural changes in the core due
to the expansion of A2. For these reasons we conclude that the hot spot B
is stationary to the limit of our ability to measure motions. For the other
jet components, the complex source structure does not permit any
determination of proper motions.

\section{Kinematics of the radio jet}
\label{section:kinematics}

\subsection{Geometry of the radio jet}
Most CSS sources show double or triple structures on kpc scales,
analogous to classical FR~I or FR~II galaxies. However, some CSS
sources show strongly asymmetric structures. At small viewing angles,
the advancing jet looks much brighter than the receding one, due to
Doppler boosting. The sidedness of radio jets can be characterized by
the jet-to-counterjet intensity ratio $R$. In VLA images
\cite{Bri95,Feng05}, 3C~48 shows two-sided structure in the north-south
direction. The southern (presumably receding) component is much weaker
than the north (advancing) one. In VLBI images (Wilkinson et al. 1991;
Worrall et al. 2004; the present paper) 3C~48 shows a one-sided jet to
the north of the nucleus. If the non-detection of the counterjet is
solely attributed to Doppler deboosting, the sideness parameter $R$ can
be estimated from the intensity ratio of jet knots to the detection
limit (derived from the $3\sigma$ off-source noise). Assuming the source is
intrinsically symmetric out to a projected separation of 600~pc (the
distance of B2 away from A1), the sideness parameter would be $>200$
for B2 and B in the 1.5-GHz image (Figure \ref{fig:vlbimap}-a).
In the highest-sensitivity image on epoch 2004C (Figure
\ref{fig:vlbimap}-d), the off-source noise in the image is 40 $\mu$Jy
beam$^{-1}$, so that the derived $R$ at component B could be as high as
$\gtrsim$900.

\begin{figure}
\includegraphics[scale=0.8]{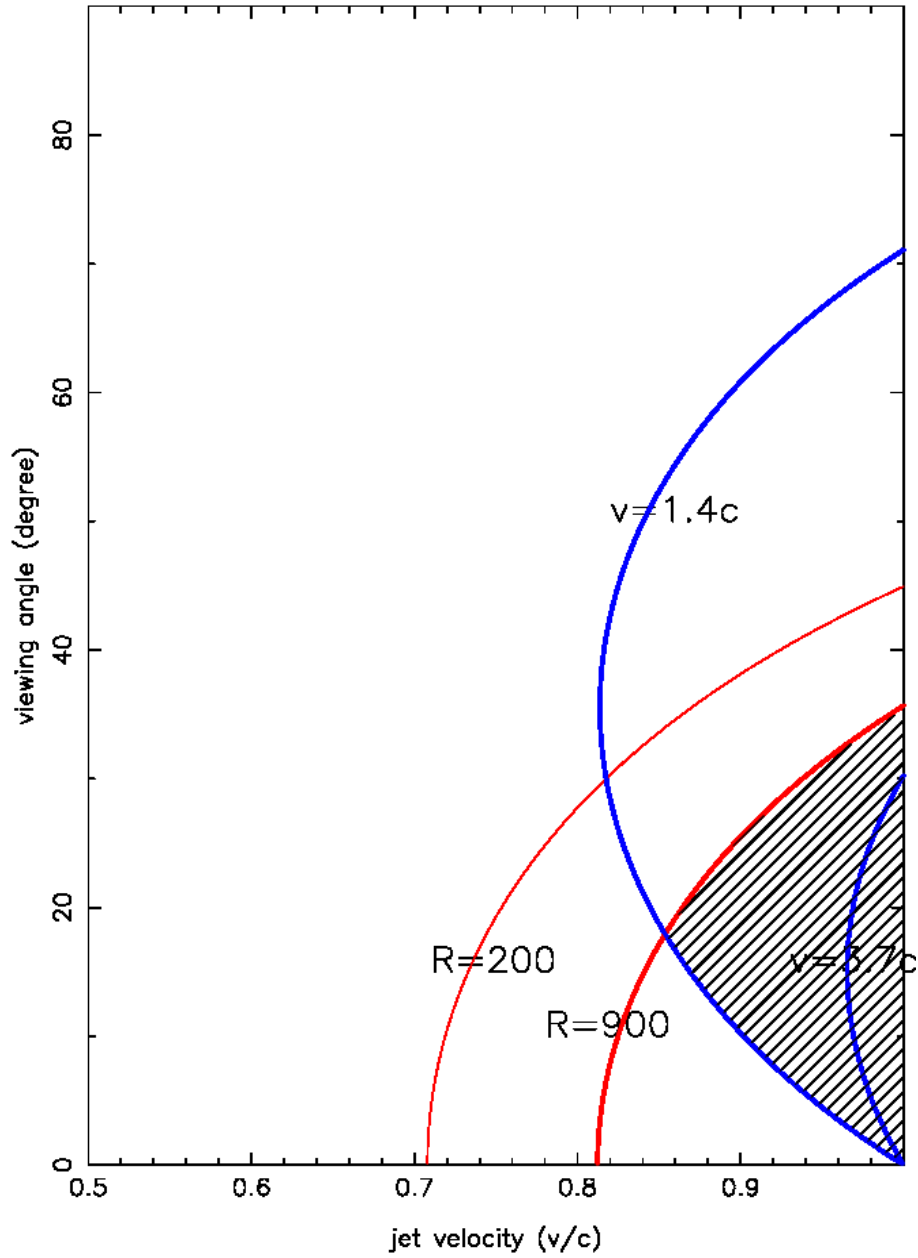}
\caption{ Constraints on source orientation and jet velocity from
the VLBI observations. The shaded region indicates the parameter
space constrained by the proper motion measurements and the
jet-to-counterjet intensity ratios.}
\label{fig:viewangle}
\end{figure}

For a smooth jet which consists of a number of unresolved components,
the jet-to-counterjet brightness ratio $R$ is related to the jet
velocity
($\beta$) and viewing angle ($\Theta$) by
$$
R=\left(\frac{1+\beta\cos\Theta}{1-\beta\cos\Theta}\right)^{2-\alpha}
$$
Assuming an optically thin spectral index $\alpha=-1.0$ for the
3C~48 jet (Figure \ref{fig:spix}), the sideness parameter $R\gtrsim900$
estimated above gives a limit of
$\beta\cos\Theta>0.81\,c$ for the projected jet velocity in
the line of sight.
Using only the combination of parameters $\beta\cos\Theta$ it is not possible
to determine the kinematics (jet speed $\beta$) and the geometry
(viewing angle $\Theta$) of the jet flow. Additional constraints may
come from the apparent transverse velocity, which is related to the
jet velocity by $\beta_{app} =
\frac{\beta\sin\Theta}{1-\beta\cos\Theta}$. In Section \ref{section:pm} we determined the
apparent velocities for components B and B2,
$\beta_{app}(B)=3.74c\pm0.35c$, $\beta_{app}(B2)=1.43c\pm0.33c$, and
so we can combine $\beta\cos\Theta$ and $\beta_{app}$ to place a constraint
on the kinematics and orientation of the outer jet.
The constraints to the jet velocity and source orientation
are shown in Figure \ref{fig:viewangle}.
The results imply that the 3C~48 jet moves at $v>0.85c$
along a viewing angle less than $35\degr$.

\subsection{Helical radio jet structure}
\label{section:helic}

As discussed in Section \ref{section:vlbimap} the bright jet knots
define a sinusoidal ridge line. This is the expected appearance of a
helically twisted jet projected on to the plane of the sky. Helical
radio jets, or jet structure with multiple bends, can be triggered by
periodic variations in the direction of ejection (e.g., precession of
the jet nozzle), and/or random perturbations at the start of the jet
(e.g., jet-cloud collisions). For example, the wiggles in the
ballistic jets in SS~433 are interpreted in terms of periodic
variation in the direction of ejection \cite{Hje81}. Alternatively,
small perturbations at the start of a coherent, smooth jet stream
might be amplified by the Kelvin-Helmholtz (K-H) instability and grow
downstream in the jet. In this case, the triggering of the helical
mode and its actual evolution in the jet are dependent on the
fluctuation properties of the initial perturbations, the dynamics of
the jet flow, and the physical properties of the surrounding
interstellar medium \cite{Har87,Har03}. In the following subsections
we consider these two models in more detail.

\subsubsection{Model 1 -- precessing jet}

\begin{figure}
\includegraphics[scale=1.3]{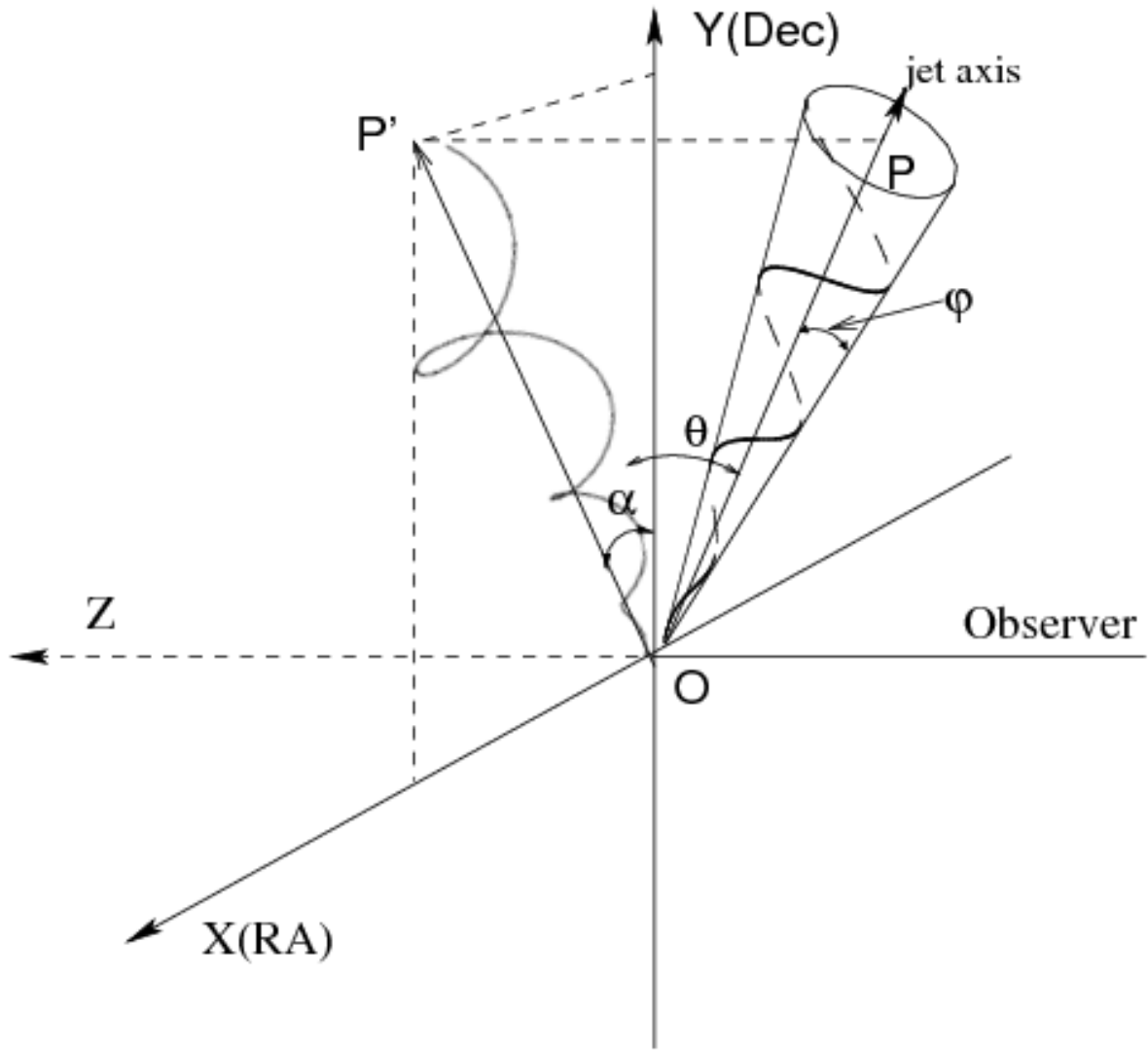}
\caption{Sketch plot of the helical jet in 3C~48. The jet knots move
on the surface of a helical cone. The XOY plane in the plot
represents the projected sky plane, and the X-axis points to the RA
direction and Y-axis to the DEC direction. The Z-axis is perpendicular
to the plane of the sky and points away from the observer.
The half of the opening angle of the helix is $\varphi$. The jet axis
'OP' is inclining by an angle of ($90-\theta$) with respect to the line
of sight. The angle $\alpha$ between the OP' and Y-axis is defined as
the position angle in the 2-Dimension CLEAN image. } \label{fig:sketch}
\end{figure}

We use a simple precession model \cite{Hje81}, taking into account
only kinematics, to model the apparently oscillatory structure of the
3C~48 radio jet. Figure \ref{fig:sketch} shows a sketch map of a 3-D
jet projected on the plane of the sky. The X- and Y-axis are defined
so that they point to the Right Ascension and Declination directions,
respectively. In the right-handed coordinate system, the Z-axis is
perpendicular to the XOY plane and the minus-Z direction points to the
observer. The jet axis is tilting toward the observer by an
inclination angle of ($90-\theta$). The observed jet axis lies at a
position angle $\alpha$. In the jet rest frame, the kinematic equation
of a precessing jet can be parameterized by jet velocity ($V_j$),
half-opening angle of the helix cone ($\varphi$) and angular velocity
(or, equivalently, precession period $P$).

To simplify the calculations, we assume a constant jet flow velocity
$V_j$, a constant opening angle $\varphi$ of the helix, and a constant
angular velocity. We ignore the width of the jet itself, so we are
actually fitting to the ridge line of the jet. The jet thickness does
not significantly affect the fitting unless it is far wider than the
opening angle of the helix cone. (We note that, although we have
measured lower proper motion velocities in B and B2 than the velocity in
the inner jet A2, this does not necessarily imply deceleration in the
outer jet flow, since the brightening at B, and to some extent at B2,
may arise mostly from stationary shocks; the proper motions of B and B2
thus represent a lower limit on the actual bulk motions of the jet.) We
further assume the origin of the precession arises from the central
black hole and accretion disk system, so that ($X_0$,$Y_0$,$Z_0$) can be
taken as zero. In the observer's frame the jet trajectory shown in the
CLEAN image can be acquired by projecting the 3-D jet on the plane of
the sky and then performing a rotation by an angle $\alpha$ in the plane
of the sky so that the Y-axis aligns to the North (Declination) and
X-axis points to the East (Right Ascension). In addition to the above
parameters, we need to define a rotation sign parameter $s_{rot}$
($s_{rot}=+1$ means counterclockwise rotation) and jet side parameter
($s_{jet}=+1$ means the jet moves toward the observer). Since we are
dealing with the advancing jet, the jet side parameter is set to 1.
Based on our calculations, we found that a clockwise rotation pattern
($s_{rot}=-1$) fits the 3C~48 jet.

To estimate the kinematical properties of the precessing jet flow, we
use the proper motion measurements of component A2 as an estimate of
the jet velocity and orientation (Figure \ref{fig:viewangle}). We have
chosen a set of parameters consistent with the curve for $V_{\rm
  app,j}=3.7c$ and a viewing angle of $17\degr$. Other combinations of
angles to the line of sight and velocities give qualitatively similar
curves. For example, if we use a lower flow speed instead, a similar
model structure can be produced by adjusting other parameters
accordingly, e.g. by increasing the precessing period by the same
factor. The high-resolution VLBI images (Figure \ref{fig:core}) show
that the innermost jet aligns to the North. So an initial position
angle $\alpha=0$ should be a reasonable estimate. The VLBI images
(Figure \ref{fig:vlbimap}) suggest that the position angle of the jet
ridge line shows an increasing trend starting from the hot spot B.
Moreover, we found that a model with a constant position angle does
not fit simultaneously to both the inner and outer jet. To simplify
the calculation, we introduced a parameter $\frac{{\rm d}\alpha}{{\rm
    d}t}$ to account for the increasing position angle in the outer
jet.

\begin{figure}
\includegraphics[width=\textwidth]{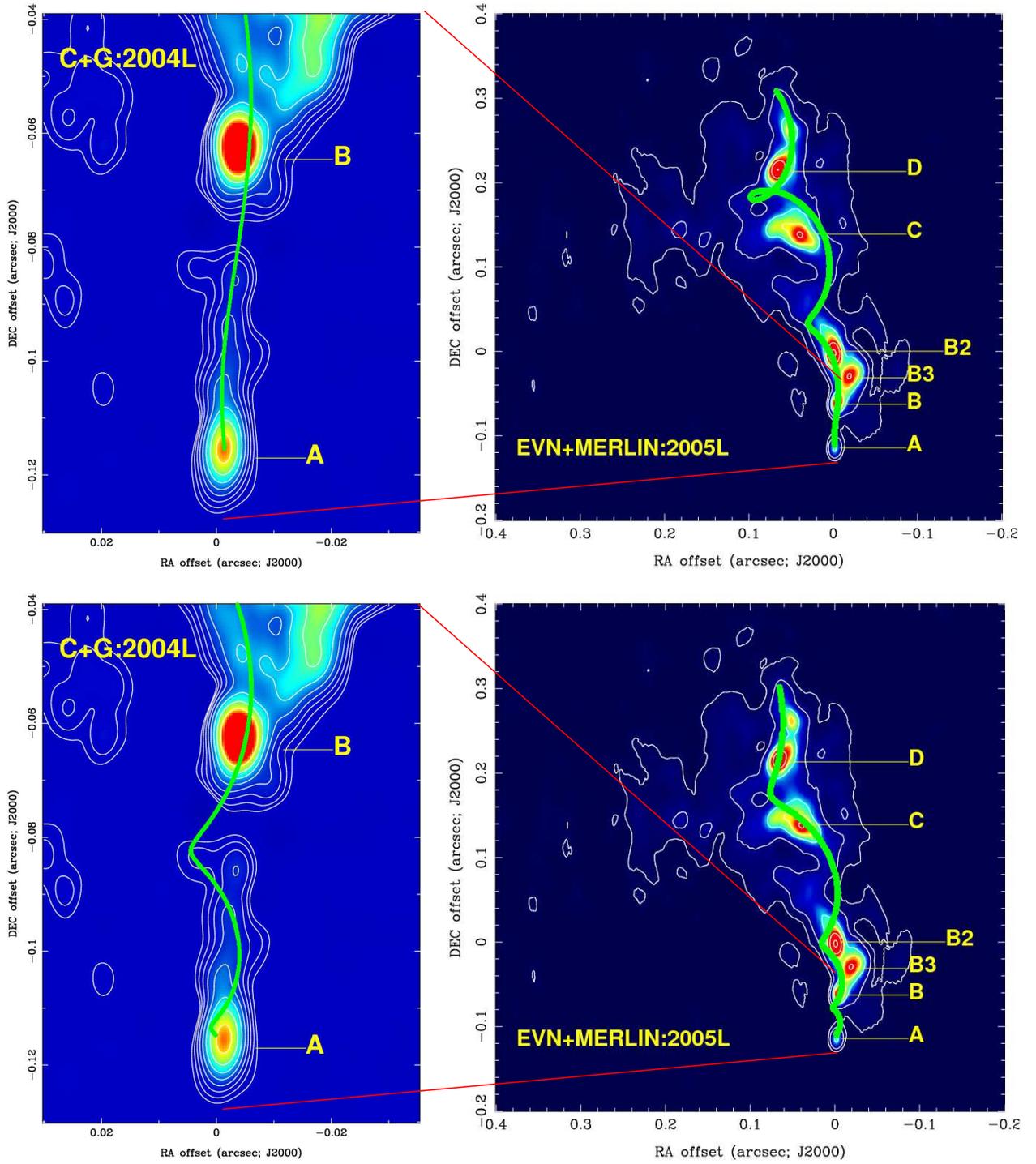}
\caption{Helical model fits overlaid on the total intensity
 images.
{\bf Upper panel} : the ridge line of the fitted precessing jet (thick
green lines) overlaid on the 1.5-GHz VLBA (left) and 1.65-GHz
EVN+MERLIN images.
The CLEAN image parameters are referred to Table \ref{tab:figpar}.
{\bf Lower panel} : the fitted jet trajectory (thick green line) from
the K-H instability model. } \label{fig:helicalfit}
\end{figure}

The fitted jet ridge line is shown (thick green line) in the upper
panel of Figure \ref{fig:helicalfit} overlaid on the total intensity
image. The assumed and fitted parameters are listed in Table
\ref{tab:helicalfit}. The modelled helix fits the general wiggling jet
structure with at least two complete periods of oscillation. The
fitted opening angle of $2.0\degr$ suggests that the line of sight
falls outside the helix cone. The initial phase angle $\phi_0$ is
loosely constrained; it is related to the reference time of the
ejection of the jet knot, $\phi_0 = 2\pi t_{ref}/P$. The fits suggest
that the reference time is $t_{ref}=-480$ yr. In the presence of the
gradual tilting of the jet axis as well as the helical coiling around
the jet axis, the fits most likely represents a superposition of the
precession of the jet knots and the nutation of the jet axis,
analogous to SS~433 (e.g. Katz et al. 1982; Begelman, King \& Pringle
2006). The fitted period of 3500 yr is then a nutation period, about 0.4
times the dynamical time scale of the jet, assuming a flow speed of
$0.965c$, while the precession period is much longer. From the rate of
the jet axis tilting, we estimate a precession period of
$\sim2\times10^5$ yr. The ratio of the estimated precession period
to the nutation period is 57:1, 2.2 times the ratio in SS~433
(which has a 162-day periodic precession and 6.3-day nodding motion:
see Begelman, King \& Pringle 2006 and references therein). The
precessing jet model predicts a smooth structure on small scales, and
a constant evolution of the wavelength so long as the jet kinetic
energy is conserved and the helix cone is not disrupted (the opening
angle of the helix cone is constant). However, the real 3C~48 jet
probably does not conserve kinetic energy, as it is characterized by a
disrupted jet and violent jet-ISM interactions. In particular, the
inner-kpc jet is seen to be physically interacting with a massive gas
system, and the observed blue-shifted NIR clouds could be driven by
the radio jet to move at velocities up to 1000 km~s$^{-1}$
\cite{Cha99,Gup05,Sto07}. The 3C~48 radio jet thus might lose a
fraction of its kinetic energy, resulting in a slowing down of the jet
flow and the shrinking of the wavelength in the outer jet, assuming
that the precessing periodicity is not destroyed.

\subsubsection{Model 2 -- Kelvin-Helmholtz instabilities}

We next investigate the interpretation of a hydrodynamic or magnetized
jet instability for a helical structure \cite{Har87,Cam86}. We used
the simple analytic model described in Steffen et al. (1995) to fit to
the helical jet trajectory in 3C~48. The kinematic equations of this
toy model are solved on the basis of the conservation of kinetic
energy $E_{kin}$ and the specific momentum in the jet motion direction
(Case 2 : Steffen et al. 1995). It is in fact identical to the
isothermal hydrodynamic model \cite{Har87} under the condition of a
small helix opening angle. Model fitting with an adiabatically
expanding jet can basically obtain similar helical twisting jet as
well, but the initial amplitude growth is much faster \cite{Har87}
than that of the isothermal jet. In this analysis we confine our
discussion to the isothermal case.

To make the calculations simple but not to lose generality, we used
similar assumptions to those of Model 1 on the jet kinematics and
geometry. (We should note that although we used an apparent velocity
$V_{\rm app,j}$ with same value in Model 1, the jet speed $V_j$ in
the K-H model is the pattern speed, and therefore the real flow speed
and the viewing angle in the K-H model are more uncertain than for the
ballistic case.) In addition, we assume that the initial perturbations
originate from a region very close to the central engine. The
calculations thus start from an initial distance of zero along the jet
axis and a small displacement $r_0$ in the rotation plane away from
the jet axis. Moreover, we assumed an initial position angle
$\alpha_0=0\degr$ , and again introduced a rate ${\rm d}\alpha/{\rm
  d}t$ to explain the eastward tilting of the jet axis. The half
opening angle, which is a parameter to be fitted, is assumed constant.
This assumption is plausible since the jet width seems not to change
much within 0.5 arcsec, indicating that the trajectory of the jet is
not disrupted even given the occurrence of a number of jet-ISM
interactions. In addition to the above morphological assumptions, the
model also assumes the conservation of specific momentum and kinetic
energy $E_{\rm kin}$ along the jet axis. The conservation of specific
momentum is equivalent to a constant velocity along the jet axis if
mass loss or entrainment are negligible. The combination of the
conservation of specific momentum and kinetic energy along the jet
axis results in a constant pitch angle along the helical jet.
Furthermore, the constant jet opening angle and pitch angle lead to a
helical geometry in which the oscillatory wavelength linearly
increases with time. The parameter $r_0$ controls how fast the
wavelength varies (Equation 12 : Steffen et al. 1995). The model
describes a self-similar helical trajectory with a number of
revolutions as long as the helical amplitude is not dampened too
rapidly.

The modelled curve is exhibited in the lower panel of Figure
\ref{fig:helicalfit}. The assumed and fitted parameters are listed in
Table \ref{tab:helicalfit}. As mentioned above, this K-H instability
model predicts that, when the helical amplitude is not dampened and the
opening angle $\varphi$ is small ($\varphi\ll
\arctan{\frac{r_0}{\lambda_0}}$), the oscillating wavelength (or period)
along the jet axis increases linearly with time. The fits give an
initial wavelength of 60 mas and initial period of 370 yr. The period
increases to $1.3\times10^4$ yr at the end of the plot window of 9000
yr. The fitted curve displays more oscillations in the inner part of the jet
and smoother structure in the outer part, due to the decreasing angular
velocity downstream. The initial transverse distance $r_0$ represents
the location where the K-H instability starts to grow in the surface
of the jet. It is associated with the varying rate of the wavelength. A
value of $r_0=1.8$ mas corresponds to a projected linear distance of 9.2
pc off the jet axis. As discussed above, the major discrepancy between the
helical model and the real 3C~48 jet could be the assumption of the
conservation of kinetic energy $E_{kin}$. We have tried to fit the helical
model without the conservation of kinetic energy but with
conserved angular momentum, which is in principle similar to Case
4 in Steffen et al. (1995). However, in this case, the modelled helix
rapidly evolves into a straight line, and thus fails to reproduce the
observed 3C~48 jet on kpc scales.

\subsubsection{Comparison of the two models}

Both two models give fits to the overall jet structure of 3C~48 within
0.45 arcsec with 2--3 complete revolutions, but they have some
differences in detail. The helical shape of the precessing jet is a
superposition of ballistic jet knots modulated by a nodding motion
(nutation). In this case, the whole jet envelope wiggles out and shows
a restricted periodicity. The observed jet structure displays a smooth shape
on rather smaller scales. If, alternatively, the coherent, smooth jet
stream is initially disturbed at the jet base, and is amplified by the
Kelvin-Helmholtz instability downstream in the jet, the jet stream
itself is bent. The resulting helical jet flow rotates faster at the
start and gradually slows down as it moves further away. If the twisted inner
jet morphology detected at 1.5 and 4.8 GHz (Figure \ref{fig:core}) is
real, this would support the K-H instability model. Further
high-dynamic-range VLBI maps of the inner jet region could test this
scenario.

In addition to the morphological discrepancy, the two models require
different physical origins. In the precessing-jet model, ballistic
knots are ejected in different directions which are associated with an
ordered rotation in the jet flow direction in the vicinity of the
central engine. If the precession results from a rotating injector at
the jet base (see discussion in Worrall et al. 2007), the precession
period of 0.2 million yr requires a radius of $17 \times
(\frac{M_{\bullet} }{10^9 M_{\odot}})^{1/3}$ pc, assuming the injector
is in a Keplerian motion around the black hole. This size scale is
much larger than the accretion disk, and so we may simply rule out the
possibility of an injection from the rotating accretion disk. Instead
the long-term precession can plausibly take place in a binary SMBH
system or a tilting accretion disk (e.g. \cite{Beg80,Lu05}). For
example, the precessing period caused by a tilting disk is $\sim
2\times10^5$ yr, assuming a $3\times10^9 M_{\sun}$ SMBH for 3C~48, a
dimensionless viscosity parameter $\alpha=0.1$ and the dimentionless
specific angular momentum of the black hole $a=0.5$ \cite{Lu05}. In
this scenario, the short-term nodding motion can then be triggered by
the tidally-induced torque on the outer brim of the wobbling accretion
disk, analogous to SS~433 \cite{Kat82,Bat00}.

On the other hand, the helical K-H instabilities modes can be
triggered by ordered or random perturbations to the jet flow. The fits
with Model 2 give an initial perturbation period $\sim370$ yr,
which leads to a radius of $\sim 0.25 \times (\frac{M_{\bullet} }{10^9
  M_{\odot}})^{1/3}$ pc where perturbations take place. This radius is
still larger than the size of the accretion disk, but at this size
scale it is still plausible for the perturbations to be due to
interactions between the jet flow and the broad-line-region clouds
(e.g. 3C120: \cite{Gom00}). However, the high Faraday depth and/or the
possible internal depolarization structure in the radio core A makes
it difficult to investigate this scenario through VLBI polarimetric
measurements. In addition, K-H instabilities would not only produce
simple helical modes, but also many other instability modes mixed
together; the K-H interpretation of the oscillatory 3C~48 jet on both
pc and kpc scales requires a selection of modes or a simple mix of
low-order modes. However, it is difficult to see how these required
modes are excited while others with higher growth rates are suppressed
(see the discussion of the wiggling filament in NGC~315 by Worrall et
al. 2007). Moreover, the K-H model does not have a ready explanation for
the observed large-scale gradual bend of the jet axis. Simple
kinematical models, such as a reflection by an oblique shock or a
pressure gradient in the Narrow-Line-Region ISM, may not be adequate to
explain the bends of the robust ($\gtrsim0.9c$) jet flow.

\section{Summary}

We have observed 3C~48 at multiple frequencies with the VLBA, EVN
and MERLIN with spatial resolutions between tens and hundreds of
parsec. Our principal results may be summarized as follows:

(1) The total-intensity MERLIN image of 3C~48 is characterized by two
components with comparable integrated flux density. A compact
component aligns with the VLBI jet, while an extended envelope
surrounds it. The extended emission structure becomes diffuse and
extends toward the northeast at $\sim$0.25 arcsec from the nucleus.
The extended component shows a steeper spectrum than the compact jet.

(2) In the VLBA and EVN images, the compact jet seen in the MERLIN
image is resolved into a series of bright knots. Knot A is
further resolved into two smaller features A1 and A2 in 4.8- and 8.3-GHz
VLBA images. A1 shows a flat spectrum with spectral index
$\alpha^{4.8}_{8.3}=-0.34\pm0.04$. A2 shows a steep spectrum with
$\alpha^{4.8}_{8.3}=-1.29\pm0.16$, and may be identified with the inner jet. The
brightness temperature of A1 is $>10^9$~K and much higher than the $T_b$
of A2. The flux densities of A1 and A2 in epoch 2004 show a 100 and 60
per cent decrease compared with those in 1996. The high brightness
temperature, flat spectrum and variability imply that A1 is the
synchrotron self-absorbed core found close to the active nucleus.

(3) Comparison of the present VLBA data with those of 1996 January
20 strongly suggests that A2 is moving, with an apparent velocity
$3.7c\pm0.4c$ to the North. Combining the apparent superluminal
motion and the jet-to-counterjet intensity ratio yields a
constraint on the jet kinematics and geometry: the jet is
relativistic ($>0.85c$) and closely aligned to the line of sight ($<35\degr$).

(4) We present for the first time VLBI polarization images of 3C~48, which
reveal polarized structures with multiple sub-components in
component C. The fractional polarization peaks at the interface between
the compact jet and the surrounding medium, perhaps consistent with a
local jet-induced shock. The systematic gradient of the EVPAs across
the jet width at C can be attributed to the combination of a gradient
in the emission-weighted intrinsic polarization angle across the jet
and possibly a systematic gradient in the RM. Changing magnetic field
directions are a possible interpretation of the RM gradient, but other
alternatives can not be ruled out. The fractional polarization of the
hot spot B increases towards higher frequencies, from $\sim1$ per cent
(1.6 GHz), $\sim2.0$ per cent (4.8 GHz) to $12$ per cent (8.3 GHz). The
relatively low degree of polarization at lower frequencies probably
results from a unresolved Faraday screen associated with the NLR clouds
and/or the internal depolarization in the jet itself. Hot spot B has a
higher RM than C, which can perhaps be attributed to a stationary shock
in the vicinity of B. The core A at all frequencies is unpolarized,
which may be the result of a tangled magnetic field in the inner part of
the jet.

(5) The combined EVN+MERLIN 1.65-GHz image and 1.5-GHz VLBA images show
that the bright knots trace out a wave-like shape within the jet. We
fitted the jet structure with a simple precession model and a K-H
instability model. Both models in general reproduce the observed
oscillatory jet trajectory, but neither of them is able to explain all
the observations. More observations are required to investigate the
physical origin of the helical pattern. Further monitoring of the proper
motion of the inner jet A2 should be able to constrain the ballistic
motion in the framework of the precessing jet. High-resolution VLBI
images of the inner jet region will be required to check whether or not
the jet flow is oscillating on scales of tens of mas, which might give a
morphological means of discriminating between the two models.
Sophisticated simulations of the jet would be needed to take into
account the deceleration of the jet flow due to kinetic energy loss via
jet-cloud interaction and radiation loss, but these are beyond the scope
of the present paper.

\section*{Acknowledgments}
TA and XYH are grateful for partial support for this work from the
National Natural Science Foundation of PR China (NSFC 10503008,
10473018) and Shanghai Natural Science Foundation (09ZR1437400). MJH
thanks the Royal Society (UK) for support. We thank Mark Birkinshaw
for helpful discussions on the jet kinematics. The VLBA is an
instrument of the National Radio Astronomy Observatory, a facility of
the US National Science Foundation operated under cooperative
agreement by Associated Universities, Inc. The European VLBI Network
(EVN) is a joint facility of European, Chinese, South African and
other radio astronomy institutes funded by their national research
councils. MERLIN is a National Facility operated by the University of
Manchester at Jodrell Bank Observatory on behalf of the UK Science and
Technology Facilities Council (STFC).

\label{lastpage}
\clearpage


\clearpage

\clearpage

\clearpage

\clearpage

\clearpage

\clearpage

\clearpage

\clearpage

\clearpage

\begin{table}
\caption{Observational parameters of 3C 48}
\begin{tabular}{c|c|c|c} \hline\hline
Array$^a$& VLBA          & EVN           & MERLIN         \\ \hline
R.A.$^b$ & 01 37 41.29943& 01 37 41.29949& 01 37 41.29675 \\
Dec.$^b$ & 33 09 35.1330 & 33 09 35.1338 & 33 09 35.5117  \\
Date     & 2004 June 25  & 2005 June 7   & 2005 June 7    \\
Time(UT) & 08:00--20:00  & 02:00--14:00  & 02:00--14:00   \\
$\tau$(hour)$^c$& 2.6/2.6/2.6   & 8.0    & 8.0            \\
Freq(GHz)$^d$   & 1.5/4.8/8.3   & 1.65   & 1.65           \\
BL(km)$^e$&20--8600      & 130--8800     & 0.3--220       \\
Calibrators& DA193, 3C138,3C345 & DA193, 3C138, 3C286 & DA193, 3C138,
3C286,
OQ208  \\
BW(MHz)  & 32           & 32             & 15             \\
Correlator& Socorro (VLBA)& JIVE (MK~IV) & Jodrell Bank   \\
\hline
\end{tabular} \\
$^a$ : Participating EVN telescopes were Jodrell Bank (Lovell 76-m), Westerbork
(phased array), Effelsberg, Onsala (25-m), Medicina, Noto, Torun, Shanghai,
Urumqi, Hartebeesthoek and Cambridge;
the MERLIN array consistsed of Defford, Cambridge, Knockin, Darnhall, MK2, Lovell and
Tabley;
all ten telescopes of the VLBA and a single VLA telescope participated in the
VLBA observations; \\
$^b$ : pointing centre of the observations; \\
$^c$ : total integrating time on 3C~48; \\
$^d$ : the central frequency of the observing band.
The VLBA observations were carried out at three frequency bands of 1.5, 5 and 8 GHz; \\
$^e$ : the projected baseline range of the array in thousands of wavelengths.
\\
\label{tab:obs}
\end{table}

\begin{table}
 \caption{Parameters of total intensity maps of Figure \ref{fig:vlbimap}}
\begin{tabular}{l|c|rrr|l|l}
\hline\hline
Label & Frequency & \multicolumn{3}{c}{Restoring Beam} & rms noise      &
Contours \\
      & (GHz)     & Maj(mas) & Min(mas) & PA(deg)      & (mJy b$^{-1}$) & (mJy
b$^{-1}$)\\ \hline
Figure 1   & 1.65 & 138 & 115 & 65.1 & 1.3  &
6.0$\times$(-2,1,2,4,8,...512)  \\
Figure 2-a & 1.51 & 8.3 & 5.3 & 1.0  & 0.25 & 1.0$\times$(1,2,4,8,16,64,256)  \\
Figure 2-b & 1.65 & 5.0 & 5.0 & 0.0  & 0.30 & 1.0$\times$(1,4,16,64,256)  \\
Figure 2-c & 4.78 & 2.7 & 1.7 & 0.0  & 0.040& 0.16$\times$(1,4,16,64,256)  \\
Figure 2-d & 8.31 & 1.8 & 1.1 & 9.7  & 0.060& 0.24$\times$(1,4,16,64,256)  \\
Figure 10$^*$& 1.65 & 16  & 10  &$-$4.4& 1.6  &
6.0$\times$(1,4,16,32,64,128)  \\
\hline
\end{tabular}\\
\label{tab:figpar}
Note: all images are registered to the phase centre of the 2005 EVN image.
$^*$: the parameters are for the 1.65-GHz EVN+MERLIN image in the right
panel.
\end{table}

\begin{table}
\caption{Properties of bright components in Figure \ref{fig:vlbimap}.}
\label{tab:modelfit}
 \begin{tabular}{llrrrrrrr}
\hline\hline
Freq.&Comp. &RA(J2000)$^a$&Dec(J2000)$^a$& S$_p$ & S$_i$  & $\theta_{maj}$ &
$\theta_{min}$ & P.A. \\
(GHz)&      &($^h$,$^m$,$^s$)&($\degr$,$\arcmin$,$\arcsec$) &(mJy b$^{-1}$) &
(mJy) & (mas) & (mas) & (degree) \\
(1) & (2) & (3) & (4) & (5) & (6) & (7) &(8) &(9)\\
\hline\hline
                                     \multicolumn{9}{c}{1996 January 20
(1996L)}
\\
1.53 &A    &01 37 41.2994260 &33 09 35.021073  & 59.94$\pm$0.84& 93.29$\pm$1.97& 4.77$\pm$0.14 & 2.58$\pm$0.18 &173.0$\pm$2.7 \\
     &B    &$-$2.69          & 53.61           &153.56$\pm$0.84&224.45$\pm$1.89& 4.02$\pm$0.06 & 2.72$\pm$0.07 &157.2$\pm$1.8 \\
     &B2   &   0.58          &113.73           & 93.20$\pm$0.78&529.51$\pm$5.17&13.46$\pm$0.13 & 8.53$\pm$0.10 & 22.8$\pm$0.8 \\
     &B3   &$-$20.97         & 84.80           & 37.04$\pm$0.78&309.78$\pm$7.23&20.38$\pm$0.45 & 8.62$\pm$0.24 &131.8$\pm$1.0 \\
     &C    &   37.56         &247.68           &24.42$\pm$0.62&385.70$\pm$10.30&27.94$\pm$0.73 &12.98$\pm$0.38 & 70.1$\pm$1.3 \\
     &D    &   70.64         &331.18           & 86.88$\pm$0.78&518.96$\pm$5.40&14.39$\pm$0.15 & 8.43$\pm$0.10 &123.7$\pm$0.8 \\
     &D2   &   50.59         &378.93           & 21.07$\pm$0.76&219.28$\pm$8.62&19.56$\pm$0.75 &11.87$\pm$0.51 & 31.4$\pm$3.1 \\
\hline
                                     \multicolumn{9}{c}{2004 June 25 (2004L)} \\

1.51 &A    &01 37 41.2993864 &33 09 35.018436  & 72.33$\pm$0.71& 99.91$\pm$1.53& 4.87$\pm$0.14&  0.00$\pm$0.00&173.2$\pm$1.1\\
     &B    &$-$2.57          &52.94            &264.06$\pm$0.72&308.76$\pm$1.38& 2.56$\pm$0.06&  1.43$\pm$0.05&  0.7$\pm$1.4\\
     &B2   &   0.73          &112.98           &130.17$\pm$0.66&657.32$\pm$3.92&13.17$\pm$0.13&  7.43$\pm$0.06& 20.6$\pm$0.4\\
     &B3   &$-$21.79         &84.88            & 47.78$\pm$0.65&371.90$\pm$5.67&18.56$\pm$0.45&  8.80$\pm$0.16&127.6$\pm$0.8\\
     &C    &   38.82         &246.04           & 28.84$\pm$0.57&376.94$\pm$7.94&28.05$\pm$0.73& 10.31$\pm$0.25& 65.4$\pm$0.8\\
     &D    &   70.82         &329.60           &119.13$\pm$0.66&616.05$\pm$3.99&13.49$\pm$0.15&  7.46$\pm$0.06&125.4$\pm$0.5\\
     &D2   &   50.52         &377.20           & 26.99$\pm$0.65&235.54$\pm$6.27&18.01$\pm$0.75& 10.54$\pm$0.31& 62.5$\pm$2.0\\

\hline
                                     \multicolumn{9}{c}{2005 June 7 (2005L)} \\

1.65 &A    &01 37 41.2993690 &33 09 35.018826  & 75.23$\pm$0.29&108.27$\pm$0.64&4.94$\pm$0.04 &1.09$\pm$0.10 &178.0$\pm$0.5 \\
     &B    &$-$2.56          & 52.92           &279.84$\pm$0.30&318.86$\pm$0.56&2.40$\pm$0.01 &1.18$\pm$0.02 & 23.7$\pm$0.6 \\
     &B2   &   2.24          &111.56          &155.31$\pm$0.27&771.75$\pm$1.59&13.44$\pm$0.03 &7.08$\pm$0.02 & 12.1$\pm$0.1 \\
     &B3   &$-$20.48         & 85.23           &50.78$\pm$0.27&297.29$\pm$1.82&17.98$\pm$0.10 &6.04$\pm$0.05 &126.9$\pm$0.2 \\
     &C    &   40.60         &249.55           &53.14$\pm$0.27&388.12$\pm$2.20&26.65$\pm$0.14 &4.51$\pm$0.05 & 62.6$\pm$0.1 \\
     &D    &   70.31         &329.77           &190.60$\pm$0.28&597.33$\pm$1.10&9.79$\pm$0.02 &5.08$\pm$0.02 &109.9$\pm$0.1 \\
     &D2   &   53.48         &377.67           &46.10$\pm$0.27&186.20$\pm$1.34&11.92$\pm$0.08 &6.00$\pm$0.06 & 40.6$\pm$0.4 \\
 \hline
                                     \multicolumn{9}{c}{1996 January 20 (1996C)}
\\
4.99 &A    &01 37 41.2993853 &33 09 35.016904 &46.80$\pm$0.04 & 60.77$\pm$0.08 &1.80$\pm$0.01 &0.43$\pm$0.01 &174.4$\pm$ 0.1 \\
     &B    &$-$2.71          & 54.85          &88.73$\pm$0.04 &135.70$\pm$0.08 &2.11$\pm$0.01 &1.37$\pm$0.01 & 36.3$\pm$ 0.2 \\
     &B2   &   0.19          &115.75          &17.99$\pm$0.03 &162.30$\pm$0.34 &9.37$\pm$0.02 &4.99$\pm$0.01 & 44.8$\pm$ 0.1 \\
     &B3   &$-$22.32         & 90.35          & 4.04$\pm$0.03 & 46.77$\pm$0.39&10.21$\pm$0.08 &6.21$\pm$0.06 &149.3$\pm$ 0.7 \\
     &D    &   71.33         &331.53          &11.75$\pm$0.02 &300.35$\pm$0.56&15.82$\pm$0.03 &9.28$\pm$0.02 &121.1$\pm$ 0.1 \\
\hline
                                     \multicolumn{9}{c}{2004 June 25 (2004C)} \\
4.78 &A1   &01 37 41.2993814 &33 09 35.016523 & 26.75$\pm$0.05 & 28.80$\pm$0.10& 1.00$\pm$0.02 &0.56$\pm$0.02 &176.3$\pm$ 2.2 \\
     &A2   &$-$0.28          &  3.39          & 16.69$\pm$0.05 & 22.86$\pm$0.11& 2.97$\pm$0.02 &0.57$\pm$0.03 &175.7$\pm$ 0.3 \\
     &B    &$-$2.63          & 55.19          &105.06$\pm$0.05 &140.80$\pm$0.11& 2.00$\pm$0.01 &0.96$\pm$0.01 & 39.8$\pm$ 0.2 \\
     &B2   &$-$0.02          &116.06          & 23.11$\pm$0.05 &161.84$\pm$0.37& 9.12$\pm$0.02 &4.61$\pm$0.02 & 42.9$\pm$ 0.2 \\
     &B3   &$-$22.39         & 90.01          &  4.07$\pm$0.04 & 52.92$\pm$0.57&12.42$\pm$0.13 &6.92$\pm$0.09 &112.9$\pm$ 0.8 \\
     &D    &   71.39         &331.59          & 14.88$\pm$0.03 &322.44$\pm$0.72&16.36$\pm$0.04 &9.22$\pm$0.03 &121.1$\pm$ 0.2 \\
 \hline
                                      \multicolumn{9}{c}{1996 January 20
(1996X)} \\
8.41 &A1   &01 37 41.2993871&33 09 35.016521 & 30.55$\pm$0.06 &  47.91$\pm$0.15& 1.39$\pm$0.01 &0.22$\pm$0.01&176.0$\pm$0.2 \\
     &A2   &$-$0.14         &  2.03          & 11.30$\pm$0.06 &  18.27$\pm$0.15& 1.39$\pm$0.03 &0.34$\pm$0.02&174.0$\pm$0.5 \\
     &B    &$-$2.73         & 55.20          & 46.00$\pm$0.06 &  89.00$\pm$0.17& 1.38$\pm$0.01 &0.71$\pm$0.01& 46.8$\pm$0.3 \\
     &B2   &$-$0.34         &115.65          &  5.04$\pm$0.05 &  96.36$\pm$0.86& 6.08$\pm$0.06 &3.54$\pm$0.03& 46.2$\pm$0.5 \\
     &D    &   71.46        &331.98          &  4.41$\pm$0.02 & 271.57$\pm$1.46&13.10$\pm$0.07 &8.09$\pm$0.04&121.5$\pm$0.4 \\
\hline
                                      \multicolumn{9}{c}{2004 June 25 (2004X)}
\\
8.31 &A1   &01 37 41.2993822 &33 09 35.016514& 18.09$\pm$0.06 &  23.85$\pm$0.12& 1.18$\pm$0.01 &0.38$\pm$0.02&178.7$\pm$0.6 \\
     &A2   &$-$0.29         &  3.41          &  6.89$\pm$0.05 &  11.19$\pm$0.14& 1.80$\pm$0.03 &0.41$\pm$0.04&175.1$\pm$0.8 \\
     &B    &$-$2.63         & 55.24          & 55.84$\pm$0.06 &  89.38$\pm$0.14& 1.34$\pm$0.01 &0.72$\pm$0.01& 37.8$\pm$0.3 \\
     &B2   &$-$0.12         &116.13          &  6.61$\pm$0.04 & 104.47$\pm$0.74& 7.14$\pm$0.05 &3.60$\pm$0.03& 43.4$\pm$0.4 \\
     &D    &   70.99        &331.82          &  4.16$\pm$0.02 & 250.24$\pm$1.39&13.11$\pm$0.07 &7.89$\pm$0.04&120.8$\pm$0.4 \\
 \hline
                                      \multicolumn{9}{c}{1996 January 20
(1996U)} \\
15.36&A1   &01 37 41.2993868& 33 09 35.016511&  9.22$\pm$0.20 &  14.59$\pm$0.48& 1.05$\pm$0.04 & 0.44$\pm$0.06 &  11.3$\pm$3.0 \\
     &A2   &$-$0.10         &  1.76          &  5.48$\pm$0.18 &   7.50$\pm$0.44& 0.93$\pm$0.08 & 0.20$\pm$0.04 &  19.4$\pm$5.0 \\
     &B    &$-$2.72         & 55.20          & 16.27$\pm$0.17 &  27.83$\pm$0.51& 1.18$\pm$0.03 & 0.47$\pm$0.03 &  41.8$\pm$1.6 \\
     &B2   &$-$0.07         &116.01          &  1.63$\pm$0.14 &  29.50$\pm$2.65& 6.59$\pm$0.57 & 2.52$\pm$0.25 &  39.8$\pm$3.3 \\
\hline\hline
\end{tabular}\label{tab:model}
$^a$ : for individual data sets, the Right Ascension and Declination positions of
the nuclear component A (A1)
in J2000.0 coordinate frame are presented;
the relative positions of jet components are given with respect to the
nuclear component
A (A1).
\end{table}

\clearpage

\begin{table}
\caption{Brightness temperature ($T_b$) of compact VLBI components}
 \begin{tabular}{rrrr} \hline\hline
$T_b$ &  A1 & A2 & B    \\ \hline
2004C & 37.4&9.8 & 53.4 \\
1996X & 74.2&8.9 & 17.2 \\
2004X & 12.8&3.7 & 22.4 \\
1996U &  2.2&2.7 & 23.2 \\
\hline
 \end{tabular} \\
Note : T$_b$ are given in units of $10^8$ K.
\label{tab:tb}
\end{table}

\begin{table}
\caption{Parameters of helical jet models}
\begin{tabular}{c|lll|rrrrr}\hline\hline
 & \multicolumn{3}{|c|}{Assumed Parameters}     &
\multicolumn{5}{|c|}{Fitted Parameters} \\
           & $V_j$
&$90-\theta$&$\alpha_0$&$d\alpha/dt$&$\varphi$&$\psi_0$&P  &$r_0$
  \\
           &(mas/yr)& (deg)     &(deg)     &(deg/$10^3$yr) & (deg)   &(deg)   &($10^3$yr)&(mas)   \\ \hline
Model 1$^a$& 0.164  & 17        &   0.0    &   1.8    & 2.0     &50.0    &3.5       &$\cdots$\\
Model 2$^b$& 0.164  & 17        &   0.0    &   1.6    & 1.5     &50.0    &0.366     &1.8     \\
\hline
\end{tabular} \\
$^a$ : a precessing jet model; \\
$^b$ : a helical-mode K-H instability model; \\
$V_j$: in the precessing model, $V_j$ represents the flow
speed in the observer's frame, taking into account relativistic
aberration effects; in the K-H model, $V_j$ denotes the pattern velocity.
The velocity is expressed in terms of proper motion in order to agree
with the coordinates used in CLEAN images; \\
$\theta$: the angle between the jet axis and the sky plane; \\
$\alpha_0$: the initial position angle of the jet axis, measured from
north to east ; \\
$d\alpha/dt$: the rate of change of position angle with time. In
the precessing model, it gives an estimate of the angular velocity
of the precession; \\
$\varphi$: half of the opening angle of the helix cone ; \\
$\psi_0$: initial phase angle of the helical jet flow; \\
$P$: in the 'Model 1', the fitted $P$ is actually the nutation period,
see discussion in Section 5.2.1; in the K-H model, $P$ represents an
initial period for the triggered perturbations; \\
$r_0$: the initial radius where the K-H instabilities starts to grow; \\
\label{tab:helicalfit}
\end{table}

\end{document}